\documentclass[%
 superscriptaddress,
 preprint,
]{revtex4-2}

\usepackage[utf8]{inputenc}
\usepackage[english]{babel}

\usepackage[dvipsnames]{xcolor}
\usepackage{graphicx}% Include figure files
\usepackage{dcolumn}% Align table columns on decimal point
\usepackage{bm}% bold math
\usepackage{upgreek}

\usepackage{amsmath}
\usepackage{lineno}

\usepackage{multirow}
\usepackage{soul}
\usepackage{hyperref}

\makeatletter
\def\@hangfrom@section#1#2#3{\@hangfrom{#1#2}#3}%\MakeTextUppercase{#3}}%
\def\@hangfroms@section#1#2{#1#2}%\MakeTextUppercase{#2}}%
\makeatother

\begin{document}

\title{Unveiling the double-peak structure of quantum oscillations in the specific heat}

\author{Zhuo Yang}
\email{zhuo.yang@issp.u-tokyo.ac.jp}
\affiliation{Institute for Solid State Physics, The University of Tokyo, Kashiwa, Chiba, 277-8581, Japan}

\author{Beno\^{i}t Fauqu\'{e}}
\affiliation{JEIP, USR 3573 CNRS, Coll\`{e}ge de France, PSL Research University, 11, Place Marcelin Berthelot, 75231 Paris Cedex 05, France}

\author{Toshihiro Nomura}
\affiliation{Institute for Solid State Physics, The University of Tokyo, Kashiwa, Chiba, 277-8581, Japan}

\author{Takashi Shitaokoshi}
\affiliation{Institute for Solid State Physics, The University of Tokyo, Kashiwa, Chiba, 277-8581, Japan}

\author{Sunghoon Kim}
\affiliation{Department of Physics, Cornell University, Ithaca NY 14853, USA}

\author{Debanjan Chowdhury}
\affiliation{Department of Physics, Cornell University, Ithaca NY 14853, USA}

\author{Zuzana Pribulov\'{a}}
\affiliation{Centre of Low Temperature Physics, Institute of Experimental Physics, Slovak Academy of Sciences, Watsonova 47, SK-04001 Ko\v{s}ice, Slovakia}

\author{Jozef Ka\v{c}mar\v{c}\'{i}k}
\affiliation{Centre of Low Temperature Physics, Institute of Experimental Physics, Slovak Academy of Sciences, Watsonova 47, SK-04001 Ko\v{s}ice, Slovakia}

\author{Alexandre Pourret}
\affiliation{Univ. Grenoble Alpes, CEA, Grenoble INP, IRIG, Pheliqs, 38000 Grenoble, France}

\author{Georg Knebel}
\affiliation{Univ. Grenoble Alpes, CEA, Grenoble INP, IRIG, Pheliqs, 38000 Grenoble, France}

\author{Dai Aoki}
\affiliation{Institute for Materials Research, Tohoku University, Oarai, Ibaraki 311-1313, Japan}

\author{Thierry Klein}
\affiliation{Univ. Grenoble Alpes, CNRS, Institut Néel, 38000 Grenoble France}

\author{Duncan K. Maude}
\affiliation{Laboratoire National des Champs Magn\'etiques Intenses, CNRS-UGA-UPS-INSA, 143 avenue de Rangueil, 31400 Toulouse, France}

\author{Christophe Marcenat}
\affiliation{Univ. Grenoble Alpes, CEA, Grenoble INP, IRIG, Pheliqs, 38000 Grenoble, France}

\author{Yoshimitsu Kohama}
\affiliation{Institute for Solid State Physics, The University of Tokyo, Kashiwa, Chiba, 277-8581, Japan}

%\date{\today}

%\linenumbers

\begin{abstract}
Quantum oscillation phenomenon is an essential tool to understand the electronic structure of quantum matter. Here we report a systematic study of quantum oscillations in the electronic specific heat $C_{\text{el}}$ in natural graphite. We show that the crossing of a single spin Landau level and the Fermi energy give rise to a double-peak structure, in striking contrast to the single peak expected from Lifshitz-Kosevich theory.
Intriguingly, the double-peak structure is predicted by the kernel term for $C_{\text{el}}/T$ in the free electron theory.
The $C_{\text{el}}/T$ represents a spectroscopic tuning fork of width $4.8 k_BT$ which can be tuned  at will to resonance. Using a coincidence method, the double-peak structure can be used to accurately determine the Land\'e $g$-factors of quantum materials. More generally, the tuning fork can be used to reveal any peak in fermionic density of states tuned by magnetic field, such as Lifshitz transition in heavy-fermion compounds.

\end{abstract}

\maketitle

\begin{flushleft}
\section*{Introduction}
\end{flushleft}

Oscillations of physical properties of materials with magnetic field are powerful tools to reveal the electronic properties of quantum matter. They range from Aharonov-Bohm oscillations\,\cite{Aharonov1959Significance} in mesoscopic rings, which provide a direct measure of the electron coherence, to quantum oscillations which provide a sensitive and incisive probe of the Fermi surface.  In the latter case, with increasing the magnetic field, the Landau quantisation of the carrier motion gives rise to a series of quantized singularities in the density of states (DOS) that cross the Fermi level, resulting in the oscillatory behaviour of various of physical quantities, such as resistivity (Shubnikov–de Haas effect), magnetic susceptibility (de Haas–van Alphen effect), thermopower and specific heat. 

Lifshitz-Kosevich (LK) theory has been widely used to describe these oscillatory phenomena\,\cite{adams1959quantum,lifshitz1956theory,Sullivan1968}, notably to extract parameters such as the effective mass and Landé $g$-factor. Although the theory is remarkably successful in describing quantum oscillations in metals over a wide range of magnetic fields and temperatures, there is growing evidence to suggest that experiment often deviates from the predicted LK behaviour\,\cite{Datars1995PRB,sandhu1996high,Hill1997PRB,woollam1971graphite,Harrison1996PRB}.
At high magnetic fields, the oscillatory magnetoresistance\,\cite{Datars1995PRB,sandhu1996high,Hill1997PRB}, magnetization\,\cite{Harrison1996PRB} and thermopower\,\cite{woollam1971graphite} exhibit a clear departure from LK theory when the systems are pushed towards the quantum limit. It is natural to expect that a similar departure is also observed in specific heat. However, the oscillatory behaviour of the specific heat in the quantum limit has yet to be fully explored. In this respect, graphite, in which the quantum limit is reached already at fields as low as 7\,T\,\cite{brandt2012semimetals}, is almost an ideal system for this purpose.

In this study, we report the quantum oscillations of specific heat in natural graphite with temperatures down to 90\,mK. 
Intriguingly, as the field increases and the system approaches the quantum limit, a characteristic double-peak structure appears in the specific heat for magnetic fields corresponding to the expected crossing of an individual spin Landau level and the Fermi energy. This result is in striking contract to the single peak feature predicted in LK theory for the quantum oscillations of specific heat, which is widely used in the literature\,\cite{riggs2011heat,michon2019thermodynamic,Kacmarcik2018PRL} (see also Supplementary Note\,1).  The double-peak structure, which unexpectedly vanishes as $T \rightarrow 0$, occurs when a narrow Landau level crosses the thermally broadened edge of the Fermi-Dirac distribution in the vicinity of the Fermi energy. We demonstrate that the double-peak structure in the oscillatory specific heat originates from the kernel term in the detailed functional form of the free electron theory expression for the specific heat\,\cite{kittel1996introduction}. 
A quantitative understanding of the double-peak structure is achieved by the comparison of a DOS model and the Slonczewski-Weiss-McCure  (SWM) tight binding Hamiltonian for graphite\,\cite{Slonc1958,McClure1960}. 
Using graphite as an example, we demonstrate that the double-peak structure provides a new way to accurately determine the $g$-factor of charge carriers without any assumptions concerning the Landau index or Fermi energy shift, and it can also be extended to other Dirac materials which is crucial in the determination of the Berry phase. Furthermore, the double-peak structure detected here is not restricted to $C_\text{el}$ in presence of Landau quantisation. It can occur in other probes related to specific heat, such as thermal conductivity, and in any system where a fermionic sea is tuned by the magnetic field such as a Lifshitz transition or in frustrated magnetic materials with fermionic like excitations.

\newpage
\begin{flushleft}
\section*{Results}
\end{flushleft}
\subsection*{Experimental results}

\begin{figure*}[t!]
  \centering
  \renewcommand{\figurename}{Fig.}
   \includegraphics[width= 0.85\linewidth]{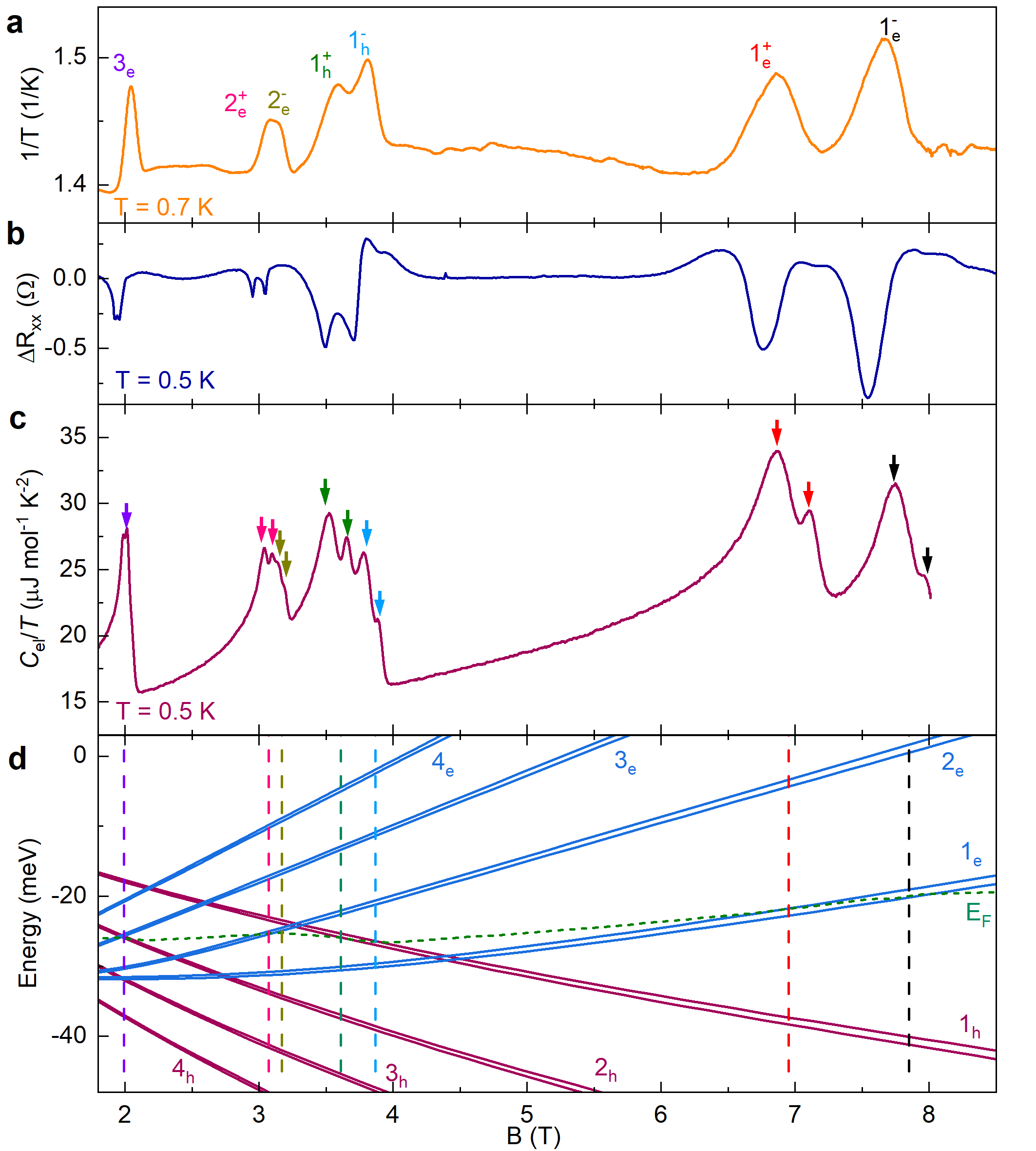}
  \caption{
  \textbf{Comparison of quantum oscillations in MCE, resistivity and specific heat.}
  \textbf{a}\,Reciprocal temperature (1/$T$) of graphite as a function of applied magnetic field in a quasi-adiabatic condition measured at initial temperature $T$ = 0.7\,K. \textbf{b} Background removed resistance $\Delta R_{xx}$ as a function of magnetic field at $T$ = 0.5\,K for natural graphite. \textbf{c} Field sweep electronic specific heat divided by temperature $C_{\text{el}}/T$ in graphite as a function of magnetic field at $T$ = 0.5\,K. \textbf{d} Electron (blue) and hole (red) Landau levels calculated within SWM-model for $B \geq$ 1.8\,T.}
  \label{fig:MCECpSWM}
\end{figure*}

When a Landau level crosses the Fermi energy, the occupation of the Landau level changes rapidly, inducing large changes in the entropy of the system, which can be probed using thermodynamic measurements. The magnetocaloric effect (MCE) measures the sample temperature as a function of applied magnetic field under quasi-adiabatic conditions. In this case, the absolute value of entropy is roughly proportional to the reciprocal sample temperature. To follow the evolution of the entropy in our graphite sample, we show in Fig.\,\ref{fig:MCECpSWM}a the measured reciprocal sample temperature ($1/T$) as a function of the magnetic field taken at an initial temperature of 0.7\,K. The field was applied along the $c$-axis of the graphite crystal for all the measurements presented in this paper. The entropy is proportional to the logarithm of the number of states within the Fermi edge, and therefore shows a maximum when a Landau level is located at the Fermi level, resulting in a series of well-defined single peaks (Supplementary Note 7) in the reciprocal sample temperature labeled as $N_{e/h}^\pm$ in Fig.\,\ref{fig:MCECpSWM}a. Here, $N$ is the Landau index, $e/h$ indicates if the Landau level originates from the electron or hole pocket, and $\pm$ indicate the spin up/down levels. For better comparison, Fig.\,\ref{fig:MCECpSWM}b shows background removed magnetoresistance $\Delta R_{xx}$ on natural graphite at 0.5\,K.

These results are in stark contrast to the electronic specific heat divided by temperature $C_{\text{el}}$/$T$ which is proportional to the temperature derivative of entropy. The $C_{\text{el}}$/$T$ of the graphite sample taken at a similar temperature ($T$ = 0.5\,K) is shown as a function of magnetic field in Fig.\,\ref{fig:MCECpSWM}c. The electronic specific heat $C_{\text{el}}$ was obtained by subtracting the phonon contribution from the total specific heat (Supplementary Note\,2). Crucially, when low-index Landau levels ($N_{e/h}<3$) cross the Fermi energy, $C_{\text{el}}$/$T$ exhibits a series of double-peak structure, as indicated by the double arrows in Fig.\,\ref{fig:MCECpSWM}c. Our observations demonstrate that as we approach the quantum limit, Landau levels crossing the Fermi energy give rise to single features in MCE and magnetoresistance, while simultaneously a novel double-peak structure is observed in the specific heat $C_{\text{el}}$/$T$. To verify that the double-peak structure in $C_\text{el}/T$ is an intrinsic effect, we have measured $C_\text{el}/T$ versus magnetic field for three different samples, together with an angle-dependence $C_\text{el}/T$ (Supplementary Note\,3). The double-peak structure in $C_\text{el}/T$ shows good reproducibility and follows the expected (for graphite) quasi-2D behavior in magnetic field, allowing us to conclude that the double-peak structure is an intrinsic effect.

\begin{figure*}[t!]
  \centering
  \renewcommand{\figurename}{Fig.}
   \includegraphics[width= 0.9\linewidth]{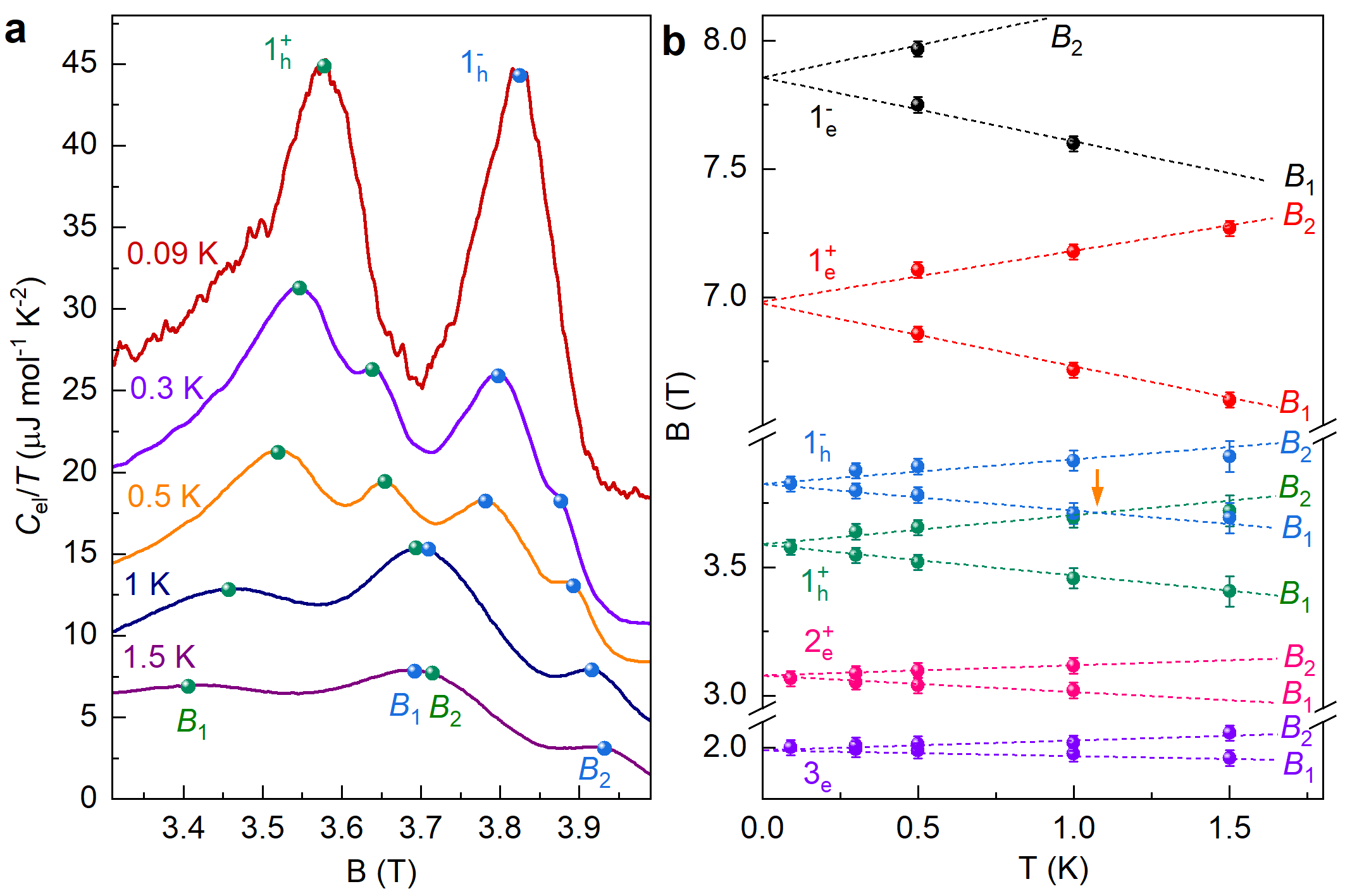}
  \caption{\textbf{Temperature dependence of the double-peak structures in $C_{\text{el}}/T$.}
  \textbf{a} Electronic specific heat divided by temperature $C_{\text{el}}/T$ in graphite as a function of magnetic field in the vicinity of the $1_h^\pm$ Landau level/Fermi energy crossing. Measurements at different temperatures are vertically offset for clarity. At higher temperature each spin Landau level gives rise to a double-peak structure (peaks indicated by symbols) as it crosses the Fermi energy. \textbf{b} Measured magnetic field position of the double-peak structure $B_1$, $B_2$ as a function of temperature for various spin up/spin down electron and hole Landau levels.
  The error bars represent one standard deviation of uncertainty.
  }
  \label{fig:TDCT1h}
\end{figure*}

Strikingly, the magnetic field splitting $\Delta B = B_2 - B_1$ of the double-peak structure in $C_{\text{el}}$/$T$ is strongly temperature dependent and vanishes as $T \rightarrow 0$.  Fig.\,\ref{fig:TDCT1h}a shows the double-peak structure in $C_{\text{el}}$/$T$ as a function of magnetic field for the $N = 1_h^{\pm}$ Landau levels measured for different temperatures from 90\,mK to 1.5\,K. The $C_{\text{el}}$/$T$ curves are vertically shifted for clarity. Symbols indicate the peak positions (corresponding to fields $B_1$ and $B_2$) for $1_h^+$ and $1_h^-$ levels, respectively. At high temperature, the splitting is clearly resolved. At lower temperatures, the splitting decreases, and the double-peak structure eventually merges into a single peak at $T$ = 90\,mK. To analyse the evolution of the splitting, we plot the magnetic field position of the double-peak structure as a function of temperature in Fig.\,\ref{fig:TDCT1h}b for various spin up/spin down hole and electron Landau levels. The peak positions $B_1$, $B_2$ scale linearly with the temperature and the extrapolated splitting vanishes at $T=0$\,K for all Landau levels. The $T$-linear dependence of the peak positions $B_1$, $B_2$ is a characteristic feature for the double-peak structure presented in this study.

The SWM Hamiltonian \cite{Slonc1958,McClure1960} with its seven tight binding parameters $\gamma_0,..., \gamma_5, \Delta$ provides a remarkably accurate description of the band structure of graphite \cite{williamson1965,Schneider2009PRL}.
In a first approach, we use the SWM-model to understand the observed Landau level crossings with the Fermi energy. The Landau levels were calculated by finding the local extrema for each Landau band ($dE_N/dk_z = 0$), where a saw-tooth-like singularity in the DOS is located. Moreover, as we approach the quantum limit, the movement of the Fermi energy to keep the charge neutrality is non-negligible, and inevitably influences the magnetic field at which a given Landau level crosses the Fermi level\,\cite{Schneider2009PRL,Schneider2010PRB,Soule1964PR}. For this reason, the Fermi level movement was calculated based on the principle of charge neutrality, that is, the difference of the electron ($n$) and hole ($p$) carrier concentration is a constant: $n-p = n_0$. For the SWM parameters, we used the values which were fine-tuned to correctly reproduce de Haas-van Alphen measurements in natural graphite \cite{Schneider2012} (Supplementary Table\,1). The calculated results are shown in Fig\,\ref{fig:MCECpSWM}d. Solid lines show the evolution of the lowest electron and hole Landau levels with magnetic field, while the green dashed line shows the calculated evolution of Fermi level. To facilitate the comparison of theory and experiment, we draw a series of vertical dashed lines in Fig\,\ref{fig:MCECpSWM}d to indicate magnetic fields corresponding to the crossing of electron/hole Landau levels with the Fermi energy. The positions of the dashed lines are in near perfect agreement with the magnetic fields of the observed peaks in MCE and the double-peak structure in $C_{\text{el}}$/$T$. 
\begin{flushleft}
\section*{Discussion}   
\end{flushleft}

\subsection*{Origin of double-peak structure}

\begin{figure*}[t!]
  \centering
  \renewcommand{\figurename}{Fig.}
   \includegraphics[width= 0.9\linewidth]{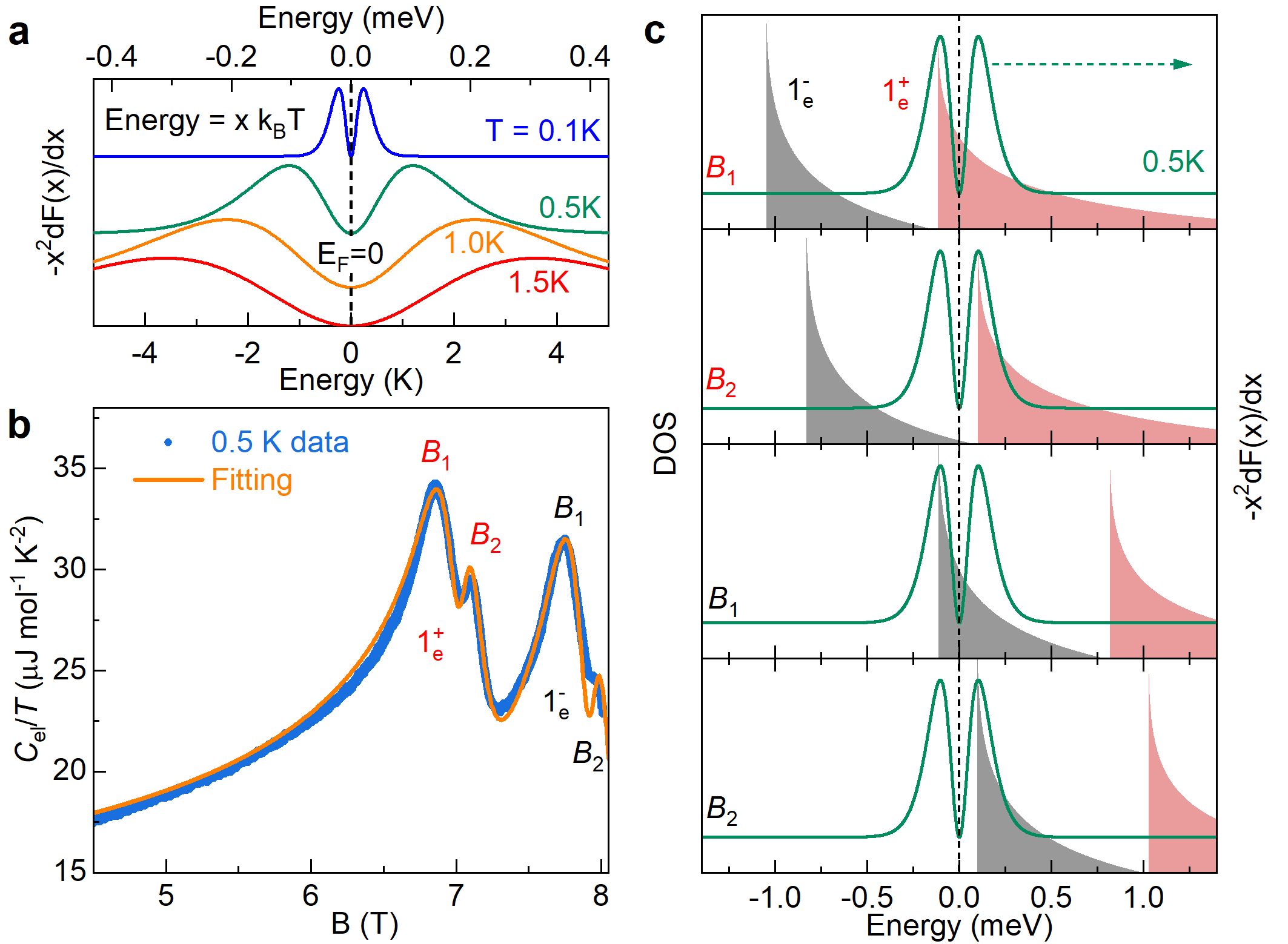}
  \caption{
    \textbf{Origin of the double-peak structure: The kernel term $-x^2 dF(x)/dx$.}
  \textbf{a} The kernel term $-x^2 dF(x)/dx$ (curves offset vertically for clarity) in the vicinity of the Fermi energy calculated at different temperatures and plotted versus $E = x k_B T$. For $-x^2 dF(x)/dx$ maxima occur at $x=\pm 2.4$ so that the splitting of the maxima is $\Delta E = 4.8 k_B T$. \textbf{b} Measured and calculated electronic specific heat divided by temperature $C_{\text{el}}$/$T$ in graphite as a function of magnetic field at $T$ = 0.5\,K. The exceptional quality of the fit, notably the position and amplitude of the double-peak structure, together with the characteristic asymmetric line-shape demonstrates the validity of our simple DOS model.  \textbf{c} Schematic to show the origin of the double-peak structure in the left panel - we plot the DOS of the $1_e^\pm$ spin split Landau level at magnetic fields corresponding to crossing the maxima in $-x^2 dF(x)/dx$ calculated here for $T=0.5$\,K.}
  \label{fig:fit}
\end{figure*}

In order to elucidate the origin of the double-peak structure in $C_{\text{el}}/T$ versus $B$, it is necessary to consider the exact form of the expression for the specific heat.  For electronic quasiparticles, $C_{\text{el}}/T$ is given by\,\cite{kittel1996introduction},
\begin{equation}
  C_{\text{el}}/T = k_B^2 \int_{-\infty}^{\infty} D(E) \left(-x^2\frac{dF(x)}{dx}\right) dx,
  \label{eqn:C/T}
\end{equation}
where $F(x)=1/(1 + e^x)$, $x = E/k_BT$ and $k_B$ is the Boltzmann constant. The specific heat depends on the convolution of the Landau level DOS $D(E)$ and a kernel term $-x^2 dF(x)/dx$ which involves the first derivative of the Fermi-Dirac distribution function. The usual approximation, removing $D(E)$ from the integral, and replacing it with $D(E_F)$, to obtain the well know formula $C_{\text{el}}=\frac{1}{3} \pi D(E_F) k_B^2T$\,\cite{kittel1996introduction}, actually suppresses the double-peak structure in $C_{\text{el}}/T$\,\cite{Marcenat2021}. As we will see, the double-peak structure in $C_{\text{el}}/T$ originates in the temperature dependent splitting of the double maxima in the kernel term $-x^2 dF(x)/dx$ (when plotted versus $E = x k_B T$). To illustrate this, in Fig.\,\ref{fig:fit}a we plot the kernel term $-x^2 dF(x)/dx$ in the vicinity of the Fermi energy ($x=0$) at different temperatures. This function shows a distinctly non-monotonic behaviour with maxima located at $x=\pm 2.4$. The maxima on either side of the Fermi energy occur at an energy $E = \pm 2.4 k_B T$ (Supplementary Note 8), so that the splitting of the maxima $\Delta E = 4.8 k_B T$ varies linearly with temperature and vanishes as $T \rightarrow 0$. Qualitatively, this exactly predicts the temperature dependence exhibited by the double-peak structure in $C_{\text{el}}/T$. The double-peak structure in quantum oscillations was also predicted in earlier theoretical calculations using an explicit expression for the specific heat\cite{zong2014thermodynamical}, however, a quantitative comparison between experiment and theory is still missing.

To obtain a quantitative comparison, we use a model DOS, calculating the specific heat  $C_{\text{el}}/T$ as the Landau level crosses the Fermi energy using Eq.\,(\ref{eqn:C/T}). The shape of Landau quantized three-dimensional DOS is saw-tooth-like, resulting from the superposition of the quantized DOS of a two dimensional system perpendicular to the field direction (delta function) and the density of states due to the dispersion along $k_z$ (DOS $\propto 1/\sqrt{E}$)\,\cite{miura2007physics}. We approximate the DOS for a single Landau level, with its ``singularity'' at $E=E_0$, by the following rigid expression for energies $E \geq E_0$,

\begin{equation}
D(E)=\frac{A}{ 1 + \sqrt{(E-E_0)/\Gamma}},
\label{eqn:DOS}
\end{equation}
The one in the denominator prevents the unphysical (in a real system) divergence of the DOS which has a maximum amplitude of $A$ at $E=E_0$. The parameter $\Gamma$ is the full width at half maximum (FWHM) of the Landau level. For simplicity all energies are calculated relative to the Fermi energy $E_F = 0$. The position of the Landau level at a given magnetic field is $E_0=(B-B_0)\,dE/dB$, $B_0$ is the magnetic field at which the Landau level crosses the Fermi energy at $T=0$. In this model, the magnetic field dependence of the Landau level energy $dE/dB$ measured relative to the Fermi energy is a fitting parameter, and thus includes the cyclotron energy and Zeeman energy, together with any movement of the Fermi energy in magnetic field. We stress that the double-peak structure, is the result of a single spin Landau level crossing the Fermi energy. However, since the spin splitting is small, the spin up/down Landau levels cross $E_F$ in quick succession, generally producing quadruple peaks. In order to locally fit the $C_{\text{el}}/T$ data we use $D_{\uparrow \downarrow}(E) = D_\uparrow(E) + D_\downarrow(E)$ with $E_0^{\uparrow \downarrow} = (B-B_0^{\uparrow \downarrow})\,dE/dB$. In this approximation the spin split Landau level rigidly shifts through the Fermi energy \emph{i.e.} the spin gap remains constant over the limited magnetic field range involved. This approximation is justified by the fact that the extracted $dE/dB$ values for a given spin up/down Landau level are quasi-identical, and for simplicity we force them to be identical in the final fit.
\begin{table}[t!]
  \renewcommand{\thetable}{\arabic{table}}
  \caption{Summary of the parameters obtained from the simple DOS model and the SWM-Hamiltonian close to where the Landau levels cross the Fermi energy. The tight binding parameters of the SWM Hamiltonian can be found in Supplementary Table 1.}
  \label{tbl_1}
  \setlength{\tabcolsep}{3mm}{
  \begin{tabular}{cccccccccc}
    \hline
     \hline
%      \cline{3-6}
    LL    &     & 1$_e^+$ & 1$_e^-$ &  2$_e^+$ &  2$_e^-$ & 3$_e$ & 1$_h^+$ & 1$_h^-$ & units\\
    \hline
    DOS & $\Gamma$  & 0.21 & 0.21   & 0.18 & 0.18 & 0.08  & 0.16 & 0.16 & meV\\
    
    & $B_0$  &  7.05 & 7.87   & 3.08 & 3.17 & 2.00 & 3.61 & 3.86 & T\\
    
     & $dE/dB$  &  1.04 &  1.04  & 4.45 & 4.45 & 7.1 & 1.74 & 1.74 & meV/T\\

    %\multirow{2}*{SWM-model}   & S$_N$  &  2.99  & 5.3 & 7.47 & 4.45  \\
    SWM   & $S_N-S_F$  &  0.98 &  0.98  & 4.06 & 4.06 & 7.47 & 2.11 & 2.11 & meV/T\\
 & S$_N$  &  2.99 &  2.99  & 5.30 & 5.30 & 7.47 & 4.45 & 4.45 & meV/T \\
       & S$_F$  &  2.01 &  2.01  & 1.24 & 1.24 & 0 & 2.34 & 2.34 & meV/T\\
    \hline
    \hline
  \end{tabular}}
\end{table}

Fig.\,\ref{fig:fit}b shows the magnetic field dependence of $C_{\text{el}}$/$T$ data in the magnetic field region where the 1$_e^{\pm}$ spin Landau levels cross the Fermi energy, together with the results of the fit. The fitting parameters used are  $\Gamma = 0.21$\,meV, $B_0^\uparrow = 7.05$\,T, $B_0^\downarrow = 7.80$\,T, and $dE/dB = 1.04$\,meV/T. 
Note that the FWHM $\Gamma$, obtained here by fitting $C_\text{el}/T$ versus $B$, is very close to the Landau level broadening $\Gamma_q = \hbar/\tau_q$ determined from the magnetic field for the onset of the Shubnikov–de Haas oscillations ($\omega_c \tau_q =1$) in natural graphite at mK temperatures reported in a previous study\,\cite{schneider2010electronic} (see Supplementary Table\,2). 
The calculated curve is in excellent agreement with the $C_{\text{el}}/T$ data, reproducing correctly the position, and the amplitude of each double-peak structure feature, together with the asymmetric line-shape, which naturally arises due to the asymmetric nature of the Landau level DOS. Fig.\,\ref{fig:fit}c schematically shows the DOS for the $1_e^\pm$ spin-split Landau level used to calculate $C_{\text{el}}/T$ at the four magnetic fields corresponding to maxima in $C_{\text{el}}/T$. For comparison, we also plot the kernel term $-x^2 dF(x)/dx$ calculated for the measurement temperature of $T=0.5$\,K. The peaks in $C_\text{el}/T$ appear at certain magnetic field when the DOS peak is tuned to the maxima of the kernel term $-x^2 dF(x)/dx$.

\begin{figure}[t!]
  \centering
  \renewcommand{\figurename}{Fig.}
   \includegraphics[width= 0.5\linewidth]{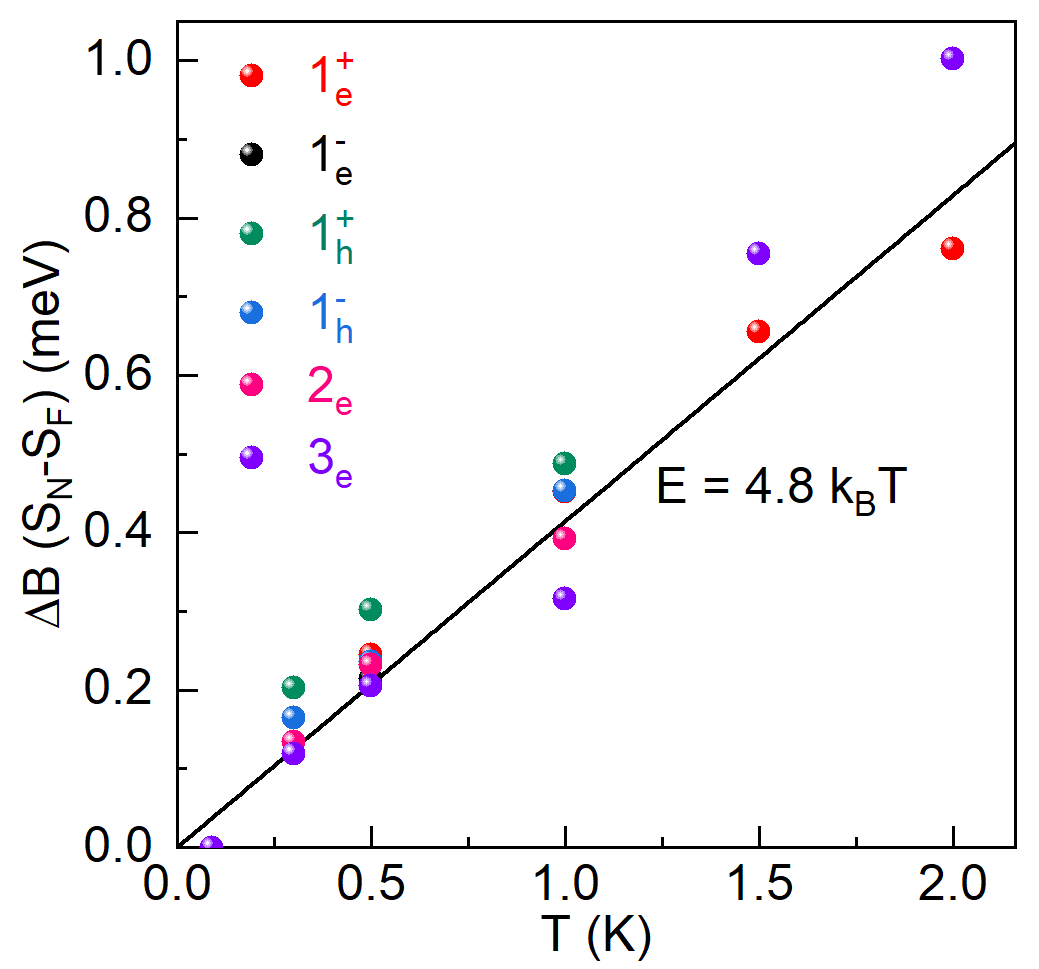}
  \caption{
   \textbf{Consistency between DOS model and SWM-Hamiltonian.}
  Temperature dependence of the energy through which the different Landau levels move $\Delta B\,(S_N -S_F)$ calculated using the movement of the Landau level relative to the Fermi energy extracted from the SWM model (see Table \ref{tbl_1}). All data collapses onto a single straight line through the origin. The solid line is the calculated $E = 4.8 k_B T$ temperature dependent splitting of the two maxima in $-x^2 dF(x)/dx$. 
  }
  \label{fig:DtPt}
\end{figure}

In Table\,\ref{tbl_1}, we summarize parameters extracted from the simple DOS model for all the Landau levels. In order to compare the values of $dE/dB$ with the predictions of the SWM Hamiltonian, we calculate the slope of SWM Landau level energy with respect to the Fermi energy in the vicinity of the crossings \emph{i.e.} $S_N - S_F$. Here, $S_N$ is the field dependence of the $N$-th SWM Landau level energy, while $S_F$ is the field dependence of the Fermi energy originated from charge neutrality condition. We see that the DOS model and SWM Landau level slopes agree to within 10\% ($dE/dB \simeq | S_N - S_F |$), which is very reasonable given the approximations involved. The good agreement between the predicted and experimental values lends further strong support to our model, and indicates that the double-peak structure in $C_{\text{el}}/T$ is a new way to access the Landau level dispersion.

A crucial test of our model for the origin of the double-peak structure is shown in Fig.\,\ref{fig:DtPt}. For each Landau level and each temperature we can compare the energy through which the Landau level moves (from field $B_1$ to $B_2$) with the energy separation of the maxima in $-x^2 dF(x)/dx$ which depends only on the temperature. The energy shift, as the magnetic field changes by $\Delta B = B_2 - B_1$, can be calculated provided we know the slope of the Landau levels (movement relative to Fermi energy). In  Fig.\,\ref{fig:DtPt} we plot the energy shift of the Landau levels $\Delta B\,(S_N -S_F)$ versus temperature $T$ using the SWM values of ($S_N-S_F$) summarized in Table\,\ref{tbl_1}. Plotted in this manner all of the data collapse onto a single straight line through the origin. The solid line is the expected splitting of the maxima in $-x^2 dF(x)/dx$, namely $E = 4.8 k_B T$. 

\subsection*{Estimate of the $dE/dB$ from the double-peak structure}

The comparison between the DOS model and SWM-model allows us to derive the following relation for a quantitive charaterization of the double-peak structure, 
\begin{equation}
  \Delta B \frac{dE}{dB} = \Delta B |S_N - S_F| = 4.8 k_B T, 
  \label{eqn:4.8kbT}
\end{equation}
Intriguingly, Eq.\,(\ref{eqn:4.8kbT}) implies that the slope of the Landau level $dE/dB$ can be estimated based on the magnetic positions $B_1$, $B_2$ of the double peaks (note that $\Delta B = B_2-B_1$). This is apparently useful for a new system with unknown shape of DOS peak, when the DOS model fitting is not applicable. Here, it is important to note that Eq.\,(\ref{eqn:4.8kbT}) is accurate provided the DOS peak is symmetric. However, in the case of asymmteric DOS peak, our simulations (Supplementary Fig.\,9) show that $\Delta B (dE/dB)$ can be $10-20$\% larger than the $4.8 k_B T$ splitting of $-x^2 dF(x)/dx$ depending on the Landau level width $\Gamma$. 
In addition, the asymmetric DOS peak also induces deviation between the peak position in $1/T$ and the center of double-peak structure (Supplementary Fig.\,10).

As $T \rightarrow 0$, we expect the double-peak structure to merge into a single peak (as seen in 90\,mK data in Fig.\,\ref{fig:TDCT1h}a), when the splitting (4.8\,$k_B T$) of maxima of kernel term $-x^2 dF(x)/dx$ is smaller than linewidth  of Landau level DOS. Due to the highly asymmetric nature of the Landau level DOS, this condition is fulfilled when the splitting of the maxima $4.8 k_B T \simeq \Gamma/2$, \emph{i.e.} half the FWHM of $D(E)$. Applying this condition, the values of $\Gamma$ extracted from the simple DOS model in Table\,\ref{tbl_1} provide a reasonable estimate of the temperature below which the double-peak structure is quenched in the experimental $C_{\text{el}}/T$ data. For example, the double-peak structure disappears between 0.3\,K and 0.09\,K for the $1_h$ Landau level in Fig.\,\ref{fig:TDCT1h}, while the predicted quench temperature $\Gamma/9.6 k_B \simeq 0.2$\,K. 

\subsection*{Estimate of the $g$-factor from the double-peak structure}
In general, to extract the $g$-factor using techniques such as SdHs, dHvA, MCE \emph{etc}, one has to know the Landau index (orbital quantum number) for each peak, and the system dependent Fermi energy shift\,\cite{shoenberg2009magnetic}.While the double-peak feature observed in specific heat allows us to estimate the $g$-factor, without having to make any assumptions concerning the Landau index or Fermi energy shift.
As a first approach, it is possible to estimate the electron and hole $g$-factors, implicitly involved in the DOS model, from the magnetic fields ($B_0^{\uparrow \downarrow}$) at which the spin Landau levels cross $E_F$. The crossing condition gives $g = 2 (dE/dB) (B_0^\downarrow - B_0^\uparrow)/\mu_B (B_0^\downarrow + B_0^\uparrow)$. Using the values in Table\,\ref{tbl_1}, we obtain $g=2.0, 2.2$ and 2.0 for the $1_e$, $2_e$ and $1_h$ Landau levels respectively. These values are close to free electron $g$-factor due to the small spin-orbit coupling of the carbon atom \cite{Dresselhaus65}, and in good agreement with electron-spin-resonance measurements in graphite \cite{Wagoner60,Kawamura83,Matsubara91,Huber04}.

It is clear that our simple DOS model provides a reasonable estimate of the $g$-factor. However, in most cases of quantum oscillations, the exact shape of the DOS is unknown, making it difficult to fit the data in order to extract the $g$-factor. Alternatively, this limitation can be overcome by using a coincidence method based on the magnetic field positions of the double peaks.
The specific heat which depends on an integral involving the kernel term $-x^2 dF(x)/dx$ represents a spectroscopic tuning fork of width $4.8 k_BT$ which can be tuned at will to resonance. For example, the observed coincidence of the $B_1$ and $B_2$ features of the $1_h^\pm$ spin Landau levels at $T=1.09$\,K and $B=3.75$\,T (marked as orange arrow in Fig.\,\ref{fig:TDCT1h}b), corresponds to the condition where the spin-split Landau levels simultaneously cross the maxima in $-x^2 dF(x)/dx$, \emph{i.e.} $g_h \mu_B B = 5.8 k_B T$ (here we use the apparent splitting in $C_{\text{el}}/T$ due to Landau level width - see Supplementary Note 10 for details) allowing us to extract the hole $g$-factor $g_h = 2.49$. Likewise, the extrapolated crossing of the $1_e^\pm$ spin Landau levels at $T=2.05$\,K and $B=7.42$\,T provides an estimate for the electron $g$-factor $g_e \mu_B B = 5.9 k_B T$ (apparent splitting in $C_{\text{el}}/T$) gives $g_e=2.42$. These values compare well with the accepted value of the electron/hole $g$-factor $g_s = 2.50$ used to fit de Haas van-Alphen data using the SWM Hamiltonian in natural graphite \cite{Schneider2012}. Note, $g$-factors measured by electron spin resonance \cite{Wagoner60,Kawamura83,Matsubara91,Huber04} are smaller ($g=2.15$) as they measure the single particle spin gap, while transport techniques measure the exchanged-enhanced spin gap. We emphasis that extracting the $g$-factor using both DOS model and coincidence method does not require the knowledge of Landau index and Fermi energy shift, which is an advantage beyond other techniques (see more discussion in Supplementary Note 14).

\subsection*{Double-peak structure in the Lifshitz transition}

\begin{table}[h!]
 \renewcommand{\thetable}{\arabic{table}}
  \caption{Summary of cyclotron, Zeeman coefficients, together with the experimentally determined energy shift $dE/dB$ from the double-peak structure, for UCoGe and CeRu$_2$Si$_2$\,\cite{aoki2014fermi,sakakibara1995absence,Knafo2012High,bastien2016lifshitz}}
  \label{tbl_LT}
  \setlength{\tabcolsep}{3mm}{
  \begin{tabular}{cccccc}
    \hline
     \hline
    compound    & $B_c$ (T) &  $\hbar e/m^*$   & $g_j m_j\mu_B$ & $dE/dB$ & Unit \\
    \hline
    UCoGe   & 9.5 &  0.008 & 0.004 &  0.37 & \text{meV/T}\\
    
    CeRu$_2$Si$_2$ & 7.7 & 0.078  & 0.109 &  1.05 & \text{meV/T}\\

    \hline
    \hline
  \end{tabular}}
\end{table}

The origin of double-peak structure reported here is not restricted to the Landau quantisation, but also applies to any system where a femionic sharp DOS peak is tuned by magnetic field. For example, the double-peak structure was observed in the vicinity of Lifshitz transition for heavy fermion compounds CeRu$_2$Si$_2$\,\cite{aoki1998thermal} and UCoGe (Supplementary Note 12-13). In graphite, $dE/dB$ of Landau levels extracted from the double-peak structure originates from the cyclotron/Zeeman energies corrected for the Fermi energy shift in magnetic field. To understand what drives the double-peak structure near the Lifshitz transition, we compare the measured $dE/dB$ with the field dependence of the cyclotron/Zeeman energies in Table\,\ref{tbl_LT}. Clearly, the $dE/dB$ values extracted from the double-peak structure in both UCoGe and CeRu$_2$Si$_2$ are too large to be explained by the Zeeman/cyclotron energy of the heavy quasiparticles ($dE/dB \gg \hbar e/m^*,\,g_jm_j\mu_B$). We conclude that the main contribution to the $dE/dB$ near the Lifshitz transition is the shift of the Fermi energy. For example, in the case of CeRu$_2$Si$_2$, the field-induced valence instability is expected to induce the large shift of the Fermi energy\,\cite{Matsuda2012Suppression}.
Therefore, the double-peak structure in $C_\text{el}/T$ can potentially be used to determine the Fermi energy shift in the vicinity of Lifshitz transition, a physical quantity that is not easy to access using other probes.

\subsection*{Kernel term for different probes}

\begin{figure}[h!]
  \centering
  \renewcommand{\figurename}{Fig.}
   \includegraphics[width= 0.5\linewidth]{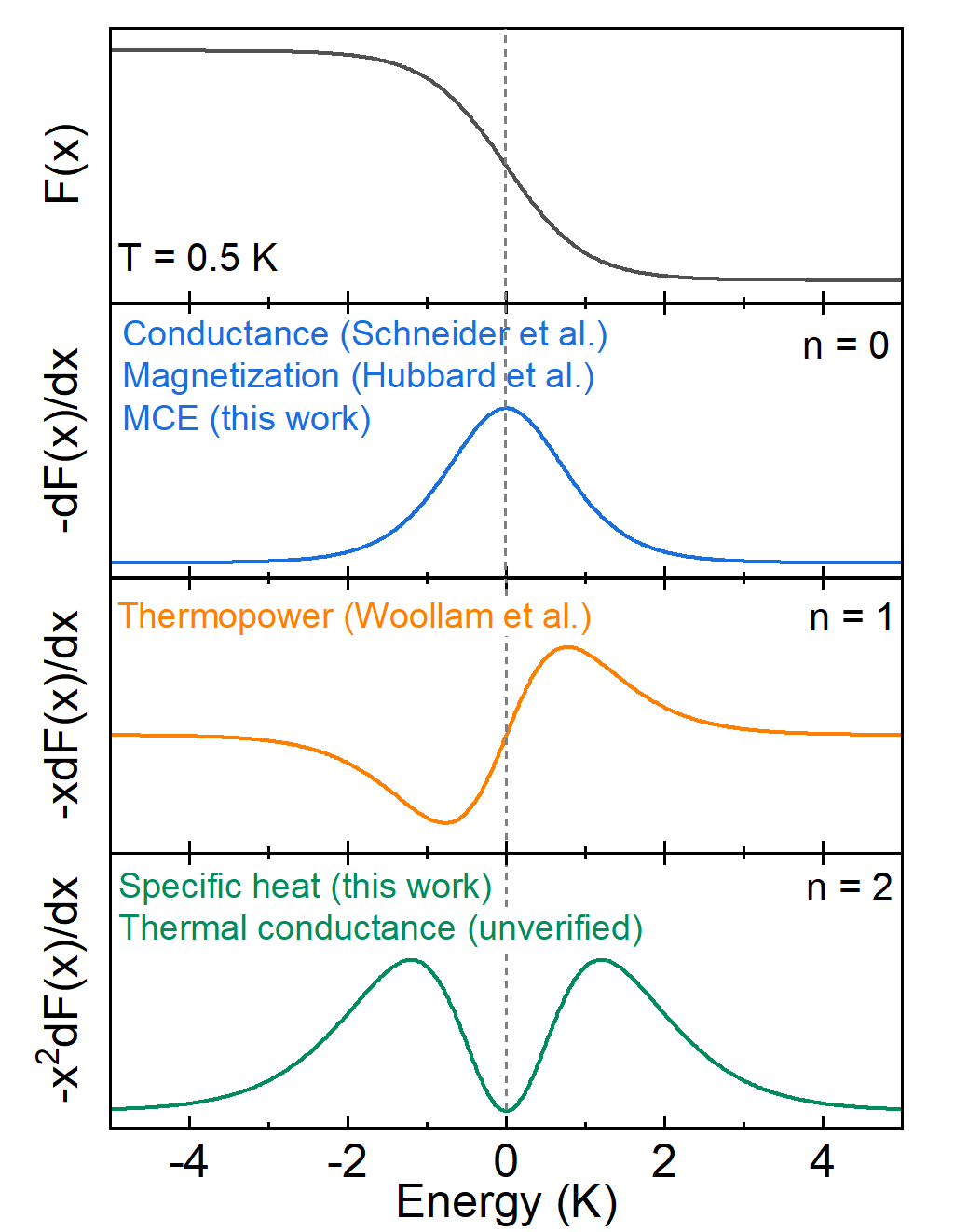}
  \caption{\textbf{Kernel terms for various thermodynamic and transport probes.}
  Fermi-Dirac distribution function $F(x)$ with $x=E/k_BT$, and the kernel terms $-x^n dF(x)/dx$ of the exact formula predicting the behaviour of different thermodynamic and transport probes. All the curves were calculated at $T$ = 0.5\,K. The features of conductance, magnetization and thermopower have been experimentally verified\,\cite{woollam1971graphite,Schneider2009PRL,hubbard2011millikelvin}. The presence of a double peak structure in thermal transport is currently a theoretical prediction.
  }
  \label{fig:Kernel}
\end{figure}

It is interesting to consider in what other thermodynamic and transport properties a double-peak structure is to be expected within the free electron theory. The exact form is an integral involving a convolution of the density of states $D(E)$ and a kernel term $-x^n dF(x)/dx$ with $n=0,1,2$ depending upon the probe considered (see Supplementary Note\,6)\,\cite{blundell2003magnetism,behnia2015fundamentals}. Fig.\,\ref{fig:Kernel} shows the three different kernel terms and the corresponding techniques. Although the shape of experimental data for different probes can be influenced by the $D(E)$ or other factors (e.g. scattering time for transport probes), it is the kernel term that determines the shape of the experimental data to be single- or double-peak feature. The single peak feature predicted for conductance and magnetization is well known\,\cite{Schneider2009PRL,hubbard2011millikelvin}. The predicted positive and negative peaks in thermopower has also been observed experimentally in graphite (see Supplementary Fig.\,7)\,\cite{woollam1971graphite,zhu2010nernst}. With the single- and double- peak features of MCE and specific heat reported in present study, the currently only unverified probe is thermal transport. As seen in Fig.\,\ref{fig:Kernel}, as thermal conductance has the same kernel term as the specific heat, simple theory predicts a similar double-peak structure in thermal transport. To the best of our knowledge, a double peak feature has yet to be observed in thermal transport, so we expect our results should stimulate further research in this direction.

We have shown that, as the quantum limit is approached in high quality graphite, the electronic specific heat divided by temperature $C_{\text{el}}/T$ exhibits a double-peak structure when a single spin Landau level crosses the Fermi energy that vanishes as $T\rightarrow 0$. A simple DOS model, combined with the predictions of the SWM Hamiltonian, successfully reproduces the double-peak structure, which can be understood with the exact form of the free electron expression for $C_{\text{el}}/T$. 
The specific heat, which depends on an integral involving the kernel term ($-x^2 dF(x)/dx$), represents a spectroscopic tuning fork of width $4.8 k_BT$ that can be tuned at will to resonance. Using a coincidence method, the double-peak structure provides a reliable estimate of the exchange enhanced $g$-factor.
Crucially, the double-peak structure is also observed in the specific heat of heavy-fermion compounds in the vicinity of the Lifshitz transition, potentially providing direct access to the Fermi energy shift at the Lifshiz transition.

\clearpage
\begin{flushleft}
\section*{Methods}  
\end{flushleft}

\subsection*{Sample description}
The measurements were performed on high quality natural graphite samples. The graphite flakes have a typical length of $\simeq 1$\,mm and thickness of $\simeq 0.1$\,mm. The weight of Sample\#1 - Sample\#3 are 0.96\,mg, 0.18\,mg, 0.23\,mg respectively.

\subsection*{Experimental setup} %\textcolor{red}{Need help from Christophe}

AC specific heat measurements were performed in a static magnetic field on natural graphite samples.
During the experiment, the specimen was attached to the backside of a bare CERNOX resistive chip by a minute amount of Apiezon grease. The resistive chip was split into heater and thermometer part by artificially making a notch along the middle line of the chip. The heater part was used to generate a periodically modulated heating power $P_{ac}$ with a frequency of 2$\omega$, which can be described as the following relation,
\begin{equation}
\renewcommand \theequation {S\arabic{equation}}
P_{ac} = \frac{R_H i^2_{ac}}{2 \omega},
\label{eqn:Pac}
\end{equation}
where $R_H$ is the resistance of heater part, $i_{ac}$ is a modulating current with a frequency of $\omega$. The induced oscillating temperature $T_{ac}$ of the sample was monitored by the thermometer part of the resistive chip. To do so, we applied a DC reading current $i_{DC}$ and monitored the induced AC voltage $V_{ac}$. Based on a precise calibration of the thermometer($R$-$T$ relation), $T_{ac}$ can be calculated from,
\begin{equation}
\renewcommand \theequation {S\arabic{equation}}
V_{ac}(2\omega) = \frac{dR_T}{dT} T_{ac}(2\omega) i_{DC},
\label{eqn:Vac}
\end{equation}
Knowing $P_{ac}$ and $T_{ac}$, specific heat can be calculated by\,\cite{kohama2010ac},
\begin{equation}
\renewcommand \theequation {S\arabic{equation}}
C = \frac{P_{ac} |\text{sin}(\phi)|}{2\omega |T_{ac}|},
\label{eqn:Cp}
\end{equation}
Here, $\phi$ stands for the phase shift between $P_{ac}$ and $T_{ac}$. By properly choosing the measurement frequency ($\omega$), $\phi$ is close to -90$^o$ ($|$sin($\phi$)$|$\,$\simeq$\,1).

To measure the angle-dependence of the specific heat in magnetic field, a CERNOX resistive chip is mounted on a copper ring attached to an attocube rotator. On the back of the copper ring, a Hall probe allows to measure the angle with the magnetic field. The misalignment between the sample and the Hall probe is estimated to be within $\pm$2 degrees.

MCE measurement was carried out in long pulsed fields with duration of 1.2\,s. The temperature of the natural graphite sample was read by the home-made RuO$_2$ thermometer, which was calibrated  in temperature and magnetic field\,\cite{imajo2021high}. The temperature of the sample was monitored and recorded during the pulse field sweeps. For both measurements, the magnetic field was applied along $c$-axis.

\subsection*{SWM Hamiltonian}

Graphite is a semi-metal with the carriers occupying a small region along the $H-K-H$ edge of the hexagonal Brillouin zone. The SWM Hamiltonian \cite{Slonc1958,McClure1960} with its seven tight binding parameters $\gamma_0,..., \gamma_5, \Delta$ provides a remarkably accurate description of the band structure of graphite \cite{williamson1965,Schneider2009PRL}. In a magnetic field, when trigonal warping is included ($\gamma_3\neq0$) levels with orbital quantum number $N$ couple to levels with orbital quantum number $N+3$ and the Hamiltonian has infinite order. Nevertheless, the infinite matrix can be truncated and numerically diagonalized, as the eigen-values converge rapidly \cite{nakao1976landau}.

The values of SWM parameters that are used in this study are shown in Supplementary Table\,1, taken from the SWM parameter set optimized to fit de Haas-van Alphen measurements in natural graphite\,\cite{Schneider2012}. They vary very little from the published values in other reports, \emph{e.g.} \cite{brandt2012semimetals,Schneider2009PRL,Schneider2010PRB}.

%\cite{Schneider2009PRL,Schneider2010PRB,Plochocka2012,Schneider2012}. 

%%%%%%%%%%%%%%%%%%%%%%%%%%%%%%%%%%%%%%%%%%%%%%%%%%%%%%%%%%%%%%%%%

\section*{Data availability}   

Source data are provided with this paper. All other data that suppot the findings of this study are available upon request to the corresponding author.

\section*{Code availability}   

The code for the SWM Hamiltonian calculation is available upon request to the corresponding author.

%\bibliography{Manuscript.bib}% Produces the bibliography via BibTeX.

\newpage

\section*{Acknowledgments}

We acknowledge the support of the LNCMI-CNRS, member of the European Magnetic Field Laboratory (EMFL).
This study has been partially supported through the EUR grant NanoX no.\ ANR-17-EURE-0009 in the framework of the ``Programme des Investissements d’Avenir''
and the Japan Society for the Promotion of Science (JSPS) KAKENHI Grants-In-Aid for Scientific Research (No. 22H00104, No. 20K14403), UTEC-UTokyo FSI Research Grant Program. B.F. is supported by JEIP-Coll\`{e}ge de France. This work is also supported by the EU H2020 project: European Microkelvin Platform (EMP), grant agreement No. 824109. Z. P. And J. K. also acknowledge the support by the EU ERDF (European regional development fund) Grant No. VA SR ITMS2014+ 313011W856 and by Slovak Scientific Grant Agency under Contract VEGA-0058/20.

\section*{Author contributions}    

Z.Y. and Y. K. conceived the study. B. F. provided high quality natural graphite samples and performed transport measurements. A. P., G. K. and D. A. grew high quality UCoGe single-crystalline sample. C. M., T. K., Z. P. and J. K. performed specific heat measurements. Y. K., T. S. and T. N. performed MCE measurements. Z. Y. and D. M. analyzed the data and performed the simulation. Z. Y., D. M., Y. K. C. M., T. K., B. F., D. C. and S. K.  discussed and interpreted the results. Z. Y., D. M. and Y. K. prepared the manuscript, with input from all other co-authors.

\section*{Competing interests}
The authors declare no competing interests.

\section*{Additional information}
\textbf{Supplementary information} accompanies this article.

\newpage

\title{Supplementary information for: Unveiling the double-peak structure of quantum oscillations in the specific heat}% Force line breaks with \\

%\linenumbers

\author{Zhuo Yang}
\email{zhuo.yang@issp.u-tokyo.ac.jp}
\affiliation{Institute for Solid State Physics, The University of Tokyo, Kashiwa, Chiba, 277-8581, Japan}

\author{Beno\^{i}t Fauqu\'{e}}
\affiliation{JEIP, USR 3573 CNRS, Coll\`{e}ge de France, PSL Research University, 11, Place Marcelin Berthelot, 75231 Paris Cedex 05, France}

\author{Toshihiro Nomura}
\affiliation{Institute for Solid State Physics, The University of Tokyo, Kashiwa, Chiba, 277-8581, Japan}

\author{Takashi Shitaokoshi}
\affiliation{Institute for Solid State Physics, The University of Tokyo, Kashiwa, Chiba, 277-8581, Japan}

\author{Sunghoon Kim}
\affiliation{Department of Physics, Cornell University, Ithaca NY 14853, USA}

\author{Debanjan Chowdhury}
\affiliation{Department of Physics, Cornell University, Ithaca NY 14853, USA}

\author{Zuzana Pribulov\'{a}}
\affiliation{Centre of Low Temperature Physics, Institute of Experimental Physics, Slovak Academy of Sciences, Watsonova 47, SK-04001 Ko\v{s}ice, Slovakia}

\author{Jozef Ka\v{c}mar\v{c}\'{i}k}
\affiliation{Centre of Low Temperature Physics, Institute of Experimental Physics, Slovak Academy of Sciences, Watsonova 47, SK-04001 Ko\v{s}ice, Slovakia}

\author{Alexandre Pourret}
\affiliation{Univ. Grenoble Alpes, CEA, Grenoble INP, IRIG, Pheliqs, 38000 Grenoble, France}

\author{Georg Knebel}
\affiliation{Univ. Grenoble Alpes, CEA, Grenoble INP, IRIG, Pheliqs, 38000 Grenoble, France}

\author{Dai Aoki}
\affiliation{Institute for Materials Research, Tohoku University, Oarai, Ibaraki 311-1313, Japan}

\author{Thierry Klein}
\affiliation{Univ. Grenoble Alpes, CNRS, Institut Néel, 38000 Grenoble France}

\author{Duncan K. Maude}
\affiliation{Laboratoire National des Champs Magn\'etiques Intenses, CNRS-UGA-UPS-INSA, 143 avenue de Rangueil, 31400 Toulouse, France}

\author{Christophe Marcenat}
\affiliation{Univ. Grenoble Alpes, CEA, Grenoble INP, IRIG, Pheliqs, 38000 Grenoble, France}

\author{Yoshimitsu Kohama}
\affiliation{Institute for Solid State Physics, The University of Tokyo, Kashiwa, Chiba, 277-8581, Japan}

%\date{\today}

\maketitle

%\tableofcontents

\makeatletter
\def\@hangfrom@section#1#2#3{\@hangfrom{#1#2}#3}%\MakeTextUppercase{#3}}%
\def\@hangfroms@section#1#2{#1#2}%\MakeTextUppercase{#2}}%
\makeatother

\begin{flushleft}
\section*{Supplementary Note 1: Deviation from LK theory in quantum oscillation of specific heat in graphite}
\end{flushleft}

\label{PiPhaseShift}

In this section, we briefly introduce the expression of quantum oscillations in specific heat predicted in LK theory, and compare it with the double-peak structure observed in experimental results.
The oscillatory component of specific heat due to Landau level quantization of the orbits is given by the extended Lifshitz-Kosevich formula\,\cite{champel2001haas},
\begin{equation}
\renewcommand \theequation {\arabic{equation}}
\Delta C_{el} (T, B) = AT\sum_{p=1}^{\infty}R_D J_0 (4\pi p \frac{t_w}{\hbar\omega_c})\text{cos}(2\pi p(\frac{\mu}{\hbar\omega_c}-\frac{1}{2}))\phi(z),
\label{eqn:L-K}
\end{equation}
where $A$ is a constant, $R_D=exp(-2\pi^2 p k_B T_D/(\hbar\omega_c))$ is the Dingle term with dingle temperature $T_D$, $J_0$ is a Bessel function of the first kind, $t_w$ is the $c$-axis hopping energy, $\mu$ is the chemical potential, $\hbar\omega_c = \hbar eB/m^*$ is cycltron energy, $\phi(z) = z(2\text{cosh}(z)/\text{sinh}^2(z)-z (1+\text{cosh}^2(z)/\text{sinh}^3(z)))$ with $z=2 \pi^2 p k_BT/(\hbar\omega_c)$.
It is important to note that $\phi(z)$ changes sign at $z = 1.6$ which induces a "$\pi$-phase shift" for $\Delta C_{el}(T,B)$. When $T$ and $B$ satisfy the condition of $z = 1.6$, the amplitude of $\Delta C_{el}(T,B)$ drops to zero, which can be used to determine the effective mass $m^*$\,\cite{riggs2011heat,bondarenko2001first,Kacmarcik2018PRL}. This kind of $\pi$ phase shift in specific heat quantum oscillations has been reported in the organic superconductor (BEDT-TTF)$_2$Cu(NCS)$_2$\,\cite{bondarenko2001first} and unusual interplay between superconductivity and field-induced charge order in YBa$_2$Cu$_3$O$_y$\,\cite{riggs2011heat,michon2019thermodynamic,Kacmarcik2018PRL}.

\begin{figure*}[t!]
  \renewcommand{\figurename}{Supplementary Figure} %\thefigure{S\arabic{figure}}
  \centering
   \includegraphics[width= 0.5\linewidth]{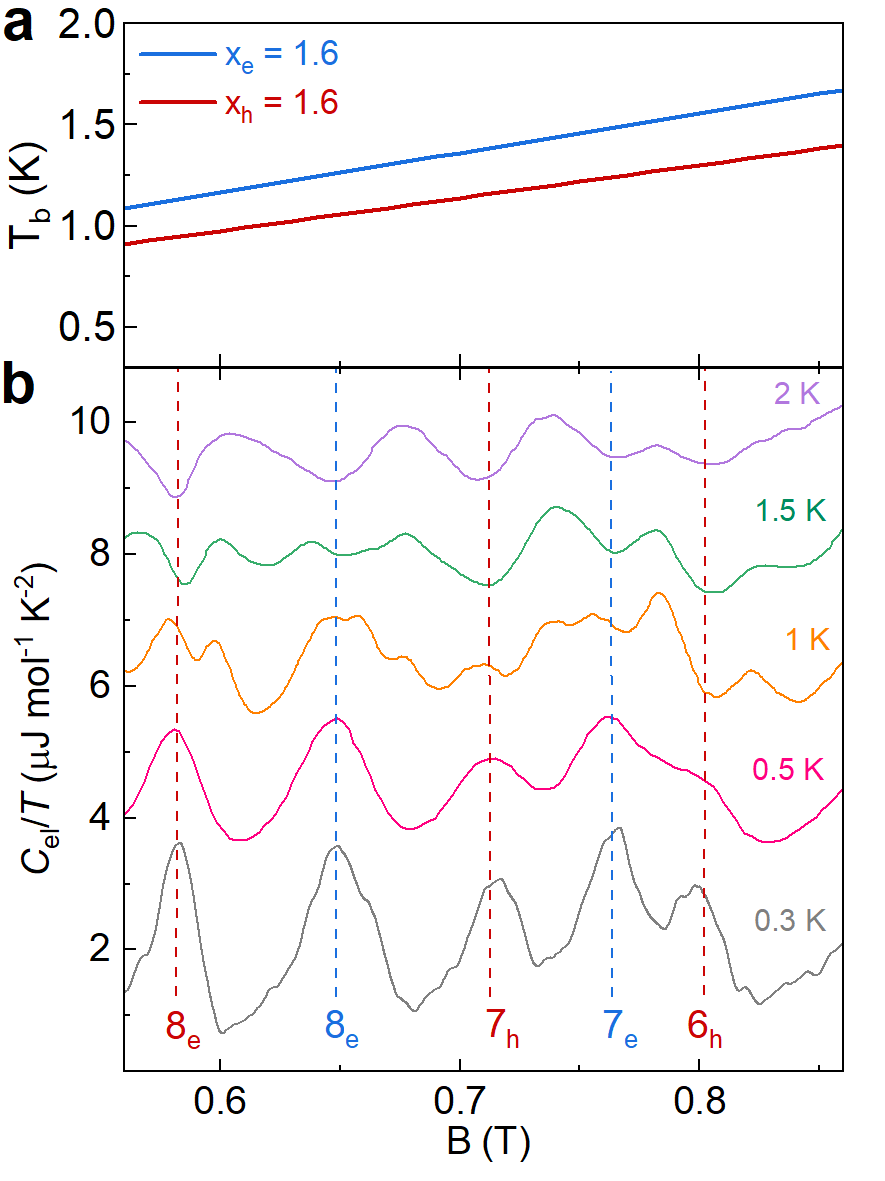} 
  \caption{\textbf{a} Calculated boundary for the $\pi$-phase shift of electron and hole pockets. \textbf{b} Oscillatory component of $C_{el}$/$T$ with high Landau index levels at indicated temperature. The dashed lines demonstrate the $\pi$-phase shift between  0.5\,K and 1.5\,K. The curves are vertically offset for clarity.}
  \label{fig:L-K}
\end{figure*}

The $\pi$-phase shift of the oscillatory components is indeed observed in graphite at low magnetic field ($B < 0.9$\,T) with Landau indexes $N>6$, which is consistent with the prediction of the extended LK formula\,\cite{Sullivan1968}. 
The quantum oscillation of specific heat in graphite is a mixture of oscillatory component from both electron and hole pockets, which make it difficult to find the zero amplitude of $\Delta C_{el}(T,B)$. Therefore, we first calculated the temperature-field boundary of the $\pi$-phase shift based on the known effective mass for electron and holes in graphite, and compare it with our experimental data. Supplementary Fig.\,\ref{fig:L-K}a shows the calculated boundary, $T_b$, for the $\pi$-phase shift ($z_{e/h} = 1.6$) using 0.056\,$m_e$ and 0.067\,$m_e$ as the electron and hole effective mass\,\cite{brandt2012semimetals}. The $\pi$-phase shift occurs around 1-1.5\,K in the field range from 0.55\,T to 0.85\,T. Supplementary Fig.\,\ref{fig:L-K}b shows the field sweep of $C_{el}/T$ taken in the temperature range of 0.3 - 2\,K. The Landau index for each oscillatory component are marked. It is clear that the $C_{el}/T$ has a $\pi$-phase shift between the 0.5\,K and 1.5\,K data, which is consistent with our calculation in Supplementary Fig.\,\ref{fig:L-K}a.

It is clear that the extended LK theory predicts a single peak feature in $C_{el}/T$ when an individual Landau level passes over the Fermi level, which has been widely used in the literature\,\cite{riggs2011heat,bondarenko2001first,michon2019thermodynamic,Kacmarcik2018PRL}. Together with the double-peak feature reported in this study, we conclude that the quantum oscillations of $C_\text{el}/T$ can appear as either single or double-peak features, depending on the width of Landau level, temperature and $dE/dB$. This suggests that care must be paid when we are trying to extract the frequency of quantum oscillations, since the frequency of the oscillations can be two times higher than the real case in the double-peak structure region.

\clearpage

\begin{flushleft}
\section*{Supplementary Note 2: Electronic specific heat of graphite}
\label{TotleSpecificHeat}
\end{flushleft}

The electronic specific heat $C_{el}$ was obtained by subtracting the phonon contribution $C_{ph}$ from the total specific heat of the specimen $C_{tot}$. At zero field and low temperature, the $C_{tot}/T$ is linearly dependent on $T^2$, and can be well described by\,\cite{kittel1996introduction},
\begin{equation}
\renewcommand \theequation {\arabic{equation}}
C_{tot}/T = \gamma + \beta T^2,
\label{eqn:gamma}
\end{equation}
where $\gamma$ is Sommerfeld coefficient, $\beta T^2$ stands for the acoustic phononic contribution to the specific heat. From the fitting of zero field specific heat data, we found $\gamma$\,=\,20$\pm$3\,$\upmu$J$\cdot$K$^{-2}\cdot$mol$^{-1}$ and $\beta$ = 28$\pm$3 $\upmu$J$\cdot$K$^{-4}\cdot$mol$^{-1}$\,\cite{Marcenat2021}. Since the phononic contribution in graphite is field independent, it is reasonable to subtract the phononic contribution for all the field range using the $\beta$ value obtained at $B$ = 0\,T.

\clearpage

\begin{flushleft}
\section*{Supplementary Note 3: Verification of intrinsic effect}
\label{SI:Intrinsic}
\end{flushleft}

In order to verify that the double-peak structure in $C_\text{el}/T$ is an intrinsic effect, we checked the reproducibility for different samples, the expected quasi-2D angle-dependence, and the reproducibility of up and down field sweeps.

\subsection*{Supplementary Note 3.1: Reproducibility of double-peak structure for different samples}

\begin{figure}[h!]
  \renewcommand{\figurename}{Supplementary Figure} %\thefigure {S\arabic{figure}}
  \centering
  \includegraphics[width= 0.8\linewidth]{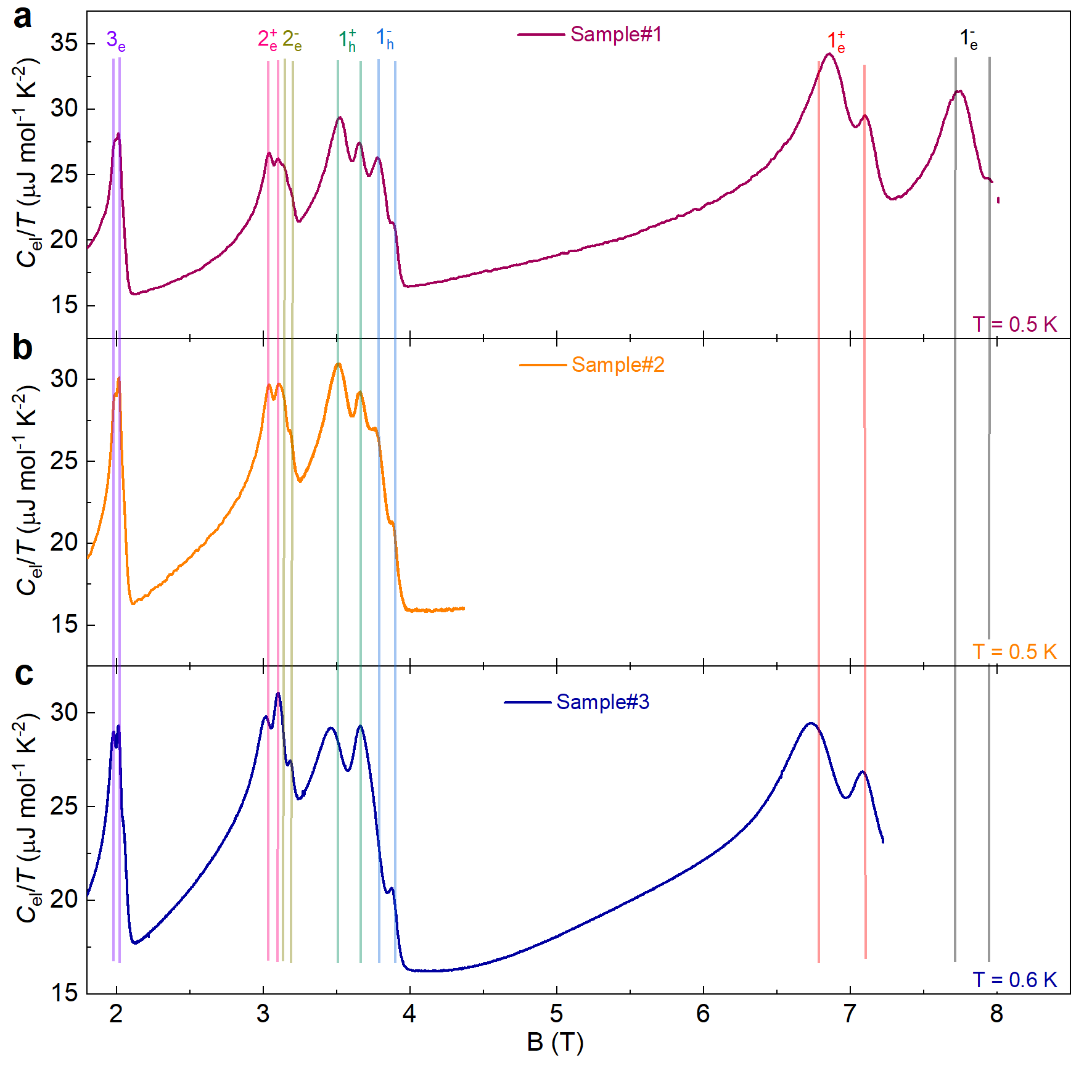}
      \caption{\textbf{a - c} Field sweep electronic
specific heat divided by temperature $C_\text{el}/T$ for Sample\#1 - Sample\#3 as a function of magnetic field. Sample\#1 and Sample\#2 were measured at $T$ = 0.5\,K. Sample\#3 was measured at $T$ = 0.6\,K. }
  \label{fig:reproducbility}
\end{figure}

Supplementary Fig.\,\ref{fig:reproducbility}a-c shows the field-sweep electronic specific heat $C_\text{el}/T$ at $T$ $\simeq$\,0.5\,K on three different natural graphite (Sample\#1 - Sample\#3). Vertical lines are guides to the eye, indicating the consistency between different sample. All three samples exhibit clear double-peak structures for each single spin-split Landau level, suggesting a good reproducbility of the double-peak structure.

\subsection*{Supplementary Note 3.2: Angle-dependence of double-peak structure}

\begin{figure}[h!]
  \renewcommand{\figurename}{Supplementary Figure} % thefigure {\arabic{figure}}
  \centering
  \includegraphics[width= 0.8\linewidth]{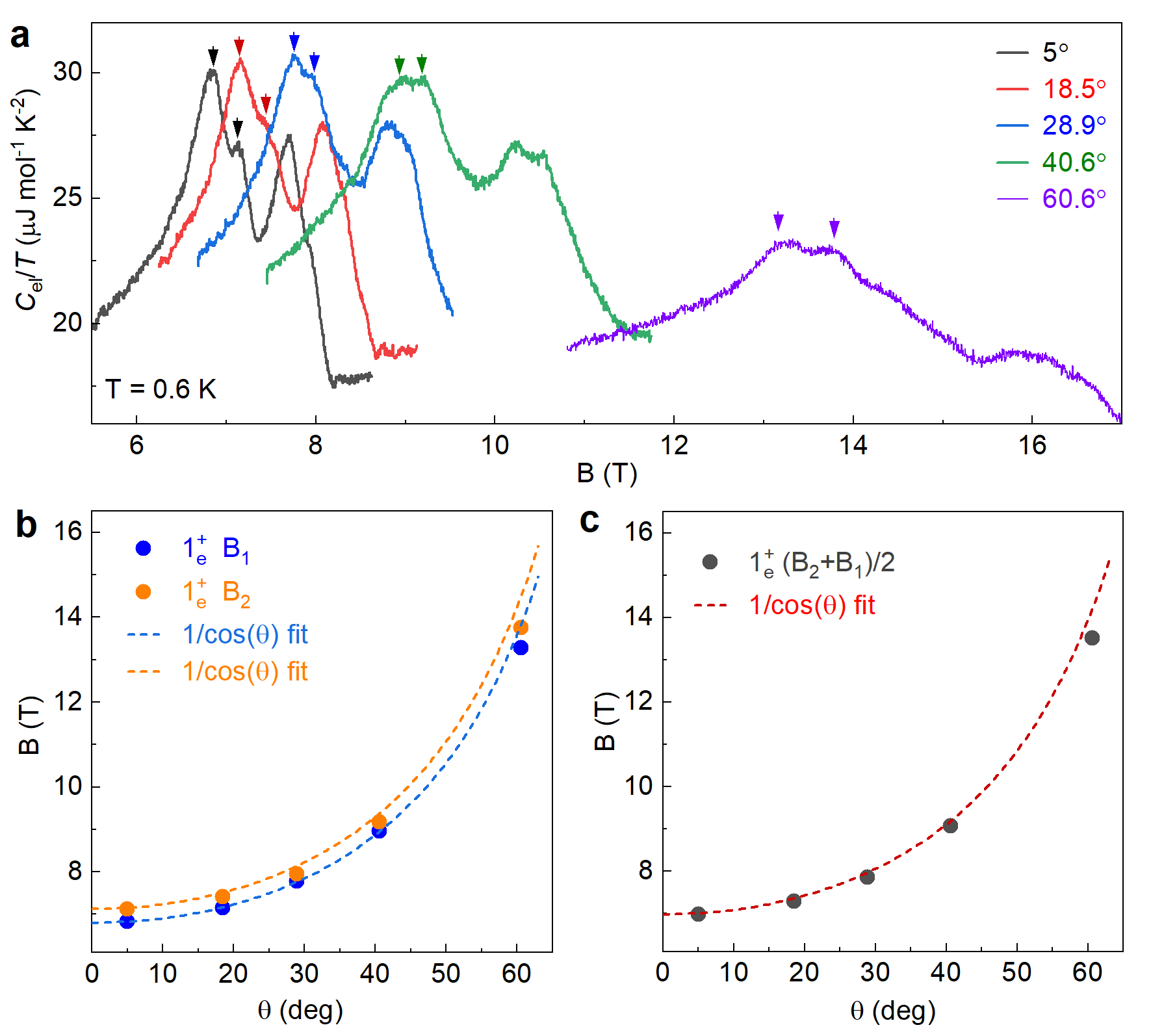}
      \caption{\textbf{a} Field sweep electronic specific heat divided by temperature $C_\text{el}/T$ of Sample\#2 measured at $T$\,=\,0.6\,K and indicated angle. \textbf{b} Angle-dependent peak position $B_1$, $B_2$ for $1_e^+$ level. \textbf{c} Angle-dependent $(B_1+B_2)/2$ for $1_e^+$ level.}
  \label{SFig:AngleDep}
\end{figure}

Graphite has a closed 3D Fermi surface, nevertheless, it shows a quasi-2D behavior in tilted magnetic fields, at least for tilt angles $\theta \leq 70^\circ$ \cite{Schneider2012}. To verify the intrinsic nature of the double peak structure, we have performed angle-dependent field sweep specific heat measurement on Sample\#2 at $T$ = 0.6\,K, as seen in Supplementary Fig.\,\ref{SFig:AngleDep}a. In this measurement, we focused on the double-peak structure for the $1_e^+$ level, where the splitting of double peaks are most clearly resolved. In Supplementary Fig.\,\ref{SFig:AngleDep}b, we show the magnetic positions of the double peaks $B_1$ $B_2$ of $1_e^+$ level as a function of angle $\theta$. Dashed lines are the fitting using $1/\text{cos}(\theta)$. The consistency between fitting and data points suggests that double-peaks follows the quasi-2D rule in magnetic field. The center of double-peaks $(B_1+B_2)/2$ also follows the quasi-2D rule, as shown in Supplementary Fig.\,\ref{SFig:AngleDep}c.

\subsection*{Supplementary Note 3.3: Reproducibility between up and down magnetic field sweeps}
\begin{figure}[h!]
  \renewcommand{\figurename}{Supplementary Figure} %  \thefigure {S\arabic{figure}}
  \centering
  \includegraphics[width= 0.9\linewidth]{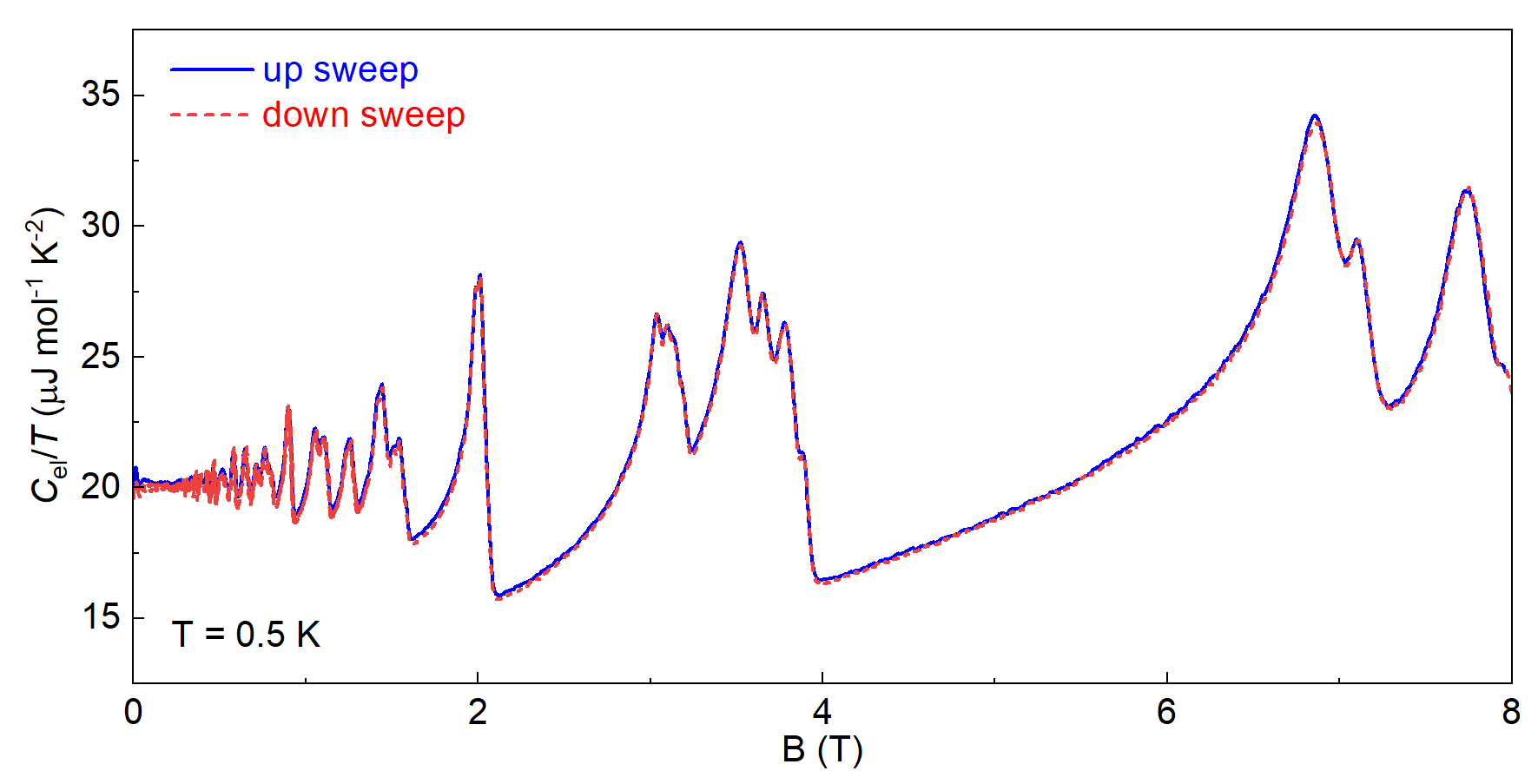}
      \caption{Field sweep electronic specific heat divided by temperature $C_\text{el}/T$ of Sample\#1 measured at $T$\,=\,0.5\,K. Red dashed and blue solid lines represent $C_\text{el}/T$ obtained in the up sweep and down sweep of the magnetic field, respectively.}
  \label{SFig:Fieldsweep}
\end{figure}

In general, the extrinsic effect originated from the experimental setup will induce discrepancy between the up sweep and down sweep field data. In Supplementary Fig.\,\ref{SFig:Fieldsweep}, we show the $C_\text{el}/T$ on Sample\#1 measured at up sweep (blue solid line) and down sweep (red dashed line) of the magnetic field. The two curves are almost identical, notably there is no hysteresis, therefore, we conclude that the double-peak structures are not extrinsic effect from the experimental setup.

\clearpage

\begin{flushleft}
\section*{Supplementary Note 4: Caculation of Landau level and Fermi energy shift within SWM-model}
\label{SWMCalculation}
\end{flushleft}

In this section, we show how the Landau levels and the movement of Fermi energy were calculated within SWM-model. Graphite is a semi-metal with the carriers occupying a small region along the $H-K-H$ edge of the hexagonal Brillouin zone. The SWM Hamiltonian \cite{Slonc1958,McClure1960} with its seven tight binding parameters $\gamma_0,..., \gamma_5, \Delta$ provides a remarkably accurate description of the band structure of graphite \cite{williamson1965,Schneider2009PRL}. In a magnetic field, when trigonal warping is included ($\gamma_3\neq0$) levels with orbital quantum number $N$ couple to levels with orbital quantum number $N+3$ and the Hamiltonian has infinite order. Nevertheless, the infinite matrix can be truncated and numerically diagonalized, as the eigen-values converge rapidly \cite{nakao1976landau}.

Supplementary Table\,\ref{SI_tbl_1} shows the value of SWM parameters that used in this study, taken from the SWM parameter set optimized to fit de Haas-van Alphen measurements in natural graphite\cite{Schneider2012}. They vary very little from the published values in other reports, \emph{e.g.} \cite{brandt2012semimetals,Schneider2009PRL,Schneider2010PRB}. 

\begin{table}[th!]
  \renewcommand{\tablename}{Supplementary Table}
 \renewcommand \thetable{\arabic{table}}
  \caption{Summary of the parameters used in the SWM tight binding Hamiltonian which are taken from \cite{Schneider2012}. The parameters $\gamma_1,...\gamma_5, \Delta$ are given in units of eV.}
  \label{SI_tbl_1}
  \begin{tabular}{ccccccccc}
    \hline
     \hline
%      \cline{3-6}
          $\gamma_0$ &  $\gamma_1$ & $\gamma_2$ & $\gamma_3$ & $\gamma_4$ & $\gamma_5$ & $\Delta$ & $g_s$ & $E_F$\,(meV)\\
    \hline
        3.15  & 0.375 & -0.0243 & 0.443 & 0.07 & 0.05 & -0.002 & 2.5 & -26.1\\
    \hline
    \hline
  \end{tabular}
 \end{table}

Under magnetic field, the band structure of graphite become quasi-one-dimensional, depending only on the wave vector along $z$-direction, which greatly simplifies our calculation. The calculated band structure along $k_z$ direction at $B$ = 2.5\,T using our SWM parameters is shown in Supplementary Fig.\,\ref{fig:SWMmodel}. In the presence of magnetic field, the bands split into Landau bands, as indicated by black solid curves. In this study, we focus on the specific heat feature for the crossing of singularity DOS and the Fermi energy. The singularity DOS locates at the local extreme of a given Landau band\,\cite{miura2007physics}, which can be found by the following relation,
\begin{equation}
\renewcommand \theequation {\arabic{equation}}
\frac{d(E_N)}{dk_z}(B)=0,
\label{eqn:LLsCal}
\end{equation}
where $E_N$ is the Landau band energy with index $N$. The local minima of electron Landau band and maxima of hole Landau band are marked as blue and red dots in Supplementary Fig.\,\ref{fig:SWMmodel}. Then, by calculating the energy of local extreme at various magnetic field, we obtained the Landau level for electrons and holes in Manuscript Fig.\,1d.

\begin{figure*}[th!]
  \renewcommand{\figurename}{Supplementary Figure} %  \thefigure {S\arabic{figure}}
  \centering
   \includegraphics[width= 0.6\linewidth]{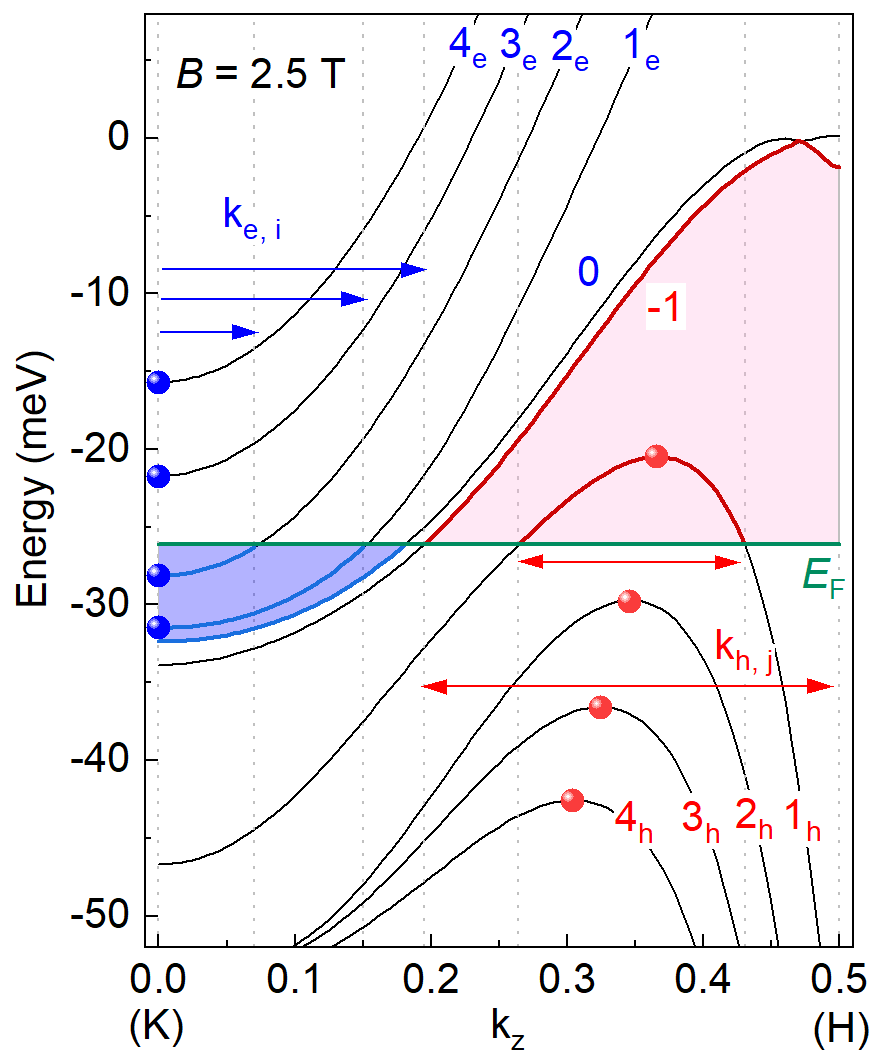}
  \caption{Band structure of graphite at $B$ = 2.5\,T calculated from SWM-model. Blue and red transparent area represent the electron and hole pockets. The local extrema for Landau levels are marked with blue (electrons) and red (holes) dots.}
  \label{fig:SWMmodel}
\end{figure*}

It has been shown in the transport measurement that the movement of Fermi energy of graphite is negligibly small at $B$ $<$ 2\,T, but becomes significant at $B$ $>$ 2\,T\,\cite{Schneider2009PRL}. The fundamental principle behind this phenomenon is the "charge neutrality" condition, that is, the difference of the electron and hole concentration should be a constant value,
\begin{equation}
\renewcommand \theequation {\arabic{equation}}
n_e - n_h = n_0,
\label{eqn:concen}
\end{equation}
where $n_e$  and $n_h$ are the electron and hole concentration, $n_0$ is a constant value representing the difference. Considering the degeneracy of the Landau bands, the electron and hole concentrations are given by\,\cite{miura2007physics},
\begin{equation}
\renewcommand \theequation {\arabic{equation}}
n_e = B\cdot\sum_i{k_{e,i}} \;\;\; and\;\;\;n_h = B\cdot\sum_j{k_{h,j}},
\label{eqn:kij}
\end{equation}
here, $k_{e,i}$ and $k_{h,j}$ are the $k_z$ distance for the occupied $i$-th electron and $j$-th hole Landau bands, as shown by blue and red arrows in Supplementary Fig.\,\ref{fig:SWMmodel}. At low magnetic field, many Landau bands are occupied, adding electrons or holes does not require significant changes of the Fermi level to fulfill the $n_e - n_h = n_0$ condition.While at high magnetic field, when only one or two Landau bands are occupied, the Fermi energy has to shift to fulfill this condition.

\clearpage

\begin{flushleft}
\section*{Supplementary Note 5: Landau level broadening $\Gamma_q$ and quantum lifetime $\tau_q$ in graphite} \label{SI:QuantumLifetime}
\end{flushleft}

In Supplementary Table\,\ref{tab:Dingle}, we summarize the quantum life times $\tau_q$ and Landau level broadening $\Gamma_q = \hbar/\tau_q$ estimated from the magnetic field for the onset of Shubnikov-de Haas oscillations ($\omega_c \tau_q = 1$) for both natural graphite (NG) and highly orientated pyrolytic graphite (HOPG) at mK temperatures\,\cite{schneider2010electronic}. The $\Gamma$ estimated from the double-peak structure in $C_{el}/T$ (0.18 - 0.21 meV) is very close to the Landau level broadening $\Gamma_q$ for NG, but much smaller than the broadening $\Gamma_q$ in HOPG. Note that our $C_\text{el}/T$ measurements were performed on natural graphite. The consistency between the $\Gamma$ extracted from $C_{el}/T$ versus $B$ and the $\Gamma_q$ extracted from  the onset of the Shubnikov-de Haas oscillations lends further support to our model.

\begin{table}[h!]
  \renewcommand{\tablename}{Supplementary Table}
  \renewcommand \thetable{\arabic{table}}
  \caption{Summary of quantum lifetime $\tau_q$ and Landau level broadening $\Gamma_q$ extracted from onset of Shubnikov-de Haas oscillations for both NG and HOPG\,\cite{schneider2010electronic}.}
  \label{tab:Dingle}
  \setlength{\tabcolsep}{3mm}{
  \begin{tabular}{ccccccc}
    \hline
     \hline
%      \cline{3-6}
        &      \multicolumn{2}{c}{NG} &  \multicolumn{2}{c}{HOPG} & \\
        &      electron & hole &  electron &  hole & unit\\
    \hline
    $\tau_q$  & 4.6 & 3.3   & 1.2 & 0.9 & ps  \\
    $\Gamma_q$ &  0.143 &  0.199  & 0.549 & 0.731 & meV\\
    \hline
    \hline
  \end{tabular}}
\end{table}

\clearpage

\begin{flushleft}
\section*{Supplementary Note 6: Exact form for various thermodynamic and transport probes}
\label{SI:ExactForm}  
\end{flushleft}

The exact forms of the charge conductance ($G$), Magnetization ($M$), Entropy ($S$, note $S \propto 1/T$ \emph{i.e.} the MCE effect), thermopower($L$), specific heat ($C$) and thermal conductance ($K$), for example when a single DOS peak $D(E)$ passes through the Fermi energy, are given by\,\cite{blundell2003magnetism,behnia2015fundamentals,kittel1996introduction},

\begin{equation}
\label{eq:conduc}
\renewcommand \theequation {\arabic{equation}}
G = 2\frac{e^2}{\hbar} \int_{-\infty}^{\infty} D(E)\left(-\frac{dF(x)}{dx}\right) dx,
\end{equation}

\begin{equation}
\renewcommand \theequation {\arabic{equation}}
M = \mu_B^2 B \int_{-\infty}^{\infty} D(E)\left(-\frac{dF(x)}{dx}\right) dx,
\end{equation}

\begin{equation}
\renewcommand \theequation {\arabic{equation}}
  S = k_B^2  \int_{-\infty}^{\infty} D(E) (- F(x)\ln{(F(x))} - (1-F(x))\ln{(1-F(x))})dx,
\end{equation}

\begin{equation}
\renewcommand \theequation {\arabic{equation}}
L = 2\frac{e^2}{\hbar}\frac{k_B}{e} \int_{-\infty}^{\infty} D(E)\left(-x\frac{dF(x)}{dx}\right) dx,
\end{equation}

\begin{equation}
\renewcommand \theequation {\arabic{equation}}
  C/T = k_B^2 \int_{-\infty}^{\infty} D(E) \left(-x^2\frac{dF(x)}{dx}\right) dx,
  \label{eqn:C/T}
\end{equation}

\begin{equation}
\label{Eq:thermalcondu}
\renewcommand \theequation {\arabic{equation}}
K/T = 2\frac{e^2}{\hbar}(\frac{k_B}{e})^2 \int_{-\infty}^{\infty} D(E)\left(-x^2\frac{dF(x)}{dx}\right) dx,
\end{equation}

\noindent Here $F(x)=(1+e^x)^{-1}$ is the Fermi-Dirac distribution function, and the dimensionless parameter $x=E/k_B T$, where the energy $E$ is measured with respect to the Fermi energy. Note, with the exception of the entropy, the thermodynamic and transport probes all depend on a kernel function of the form $-x^n dF/dx$ with $n=0,1,2$. It is the value of $n$ which decides the response of a given thermodynamic or transport probe when a DOS singularity crosses the Fermi energy. Note that, as discussed below, the derivative with respect to temperature of the entropy $dS/dT$ does depend on the kernel function $-x^n dF/dx$ with $n=2$.

\clearpage
\begin{flushleft}
\section*{Supplementary Note 7: Calculated entropy near the crossing point of a Landau level and the Fermi energy}
\label{EntropyCalculation}
\end{flushleft}

In this section, we show the calculated entropy $S$ in the vicinity of a crossing point of a Landau level with the Fermi energy in order to fully understand the magneto-caloric effect (MCE) results in the manuscript. A single Landau level DOS was constructed using Manuscript Eq.\,(2) with parameters listed in Supplementary Fig.\,\ref{fig:Entropy}a. The crossing field $B_0$ was set to 5\,T. The entropy $S_{el}$ for Fermionic quasiparticles is given by\,\cite{Hoffmann2020From},

\begin{equation}
\renewcommand \theequation {\arabic{equation}}
  S_{el} = k_B^2  \int_{-\infty}^{\infty} D(E) (- F(x)\ln{(F(x))} - (1-F(x))\ln{(1-F(x))})dx,
\label{eqn:Entropy}
\end{equation}
where $F(x)=1/(1 + e^x)$, $x = E/k_BT$ and $k_B$ is the Boltzmann constant. Taking a temperature derivative of Supplementary Eq.\,(\ref{eqn:Entropy}) results in, 
\begin{equation}
\renewcommand \theequation {\arabic{equation}}
  \frac{dS_{el}}{dT} = k_B^2 \int_{-\infty}^{\infty} D(E) x^2(-\frac{dF(x)}{dx})dx = C_{el}/T,
\label{eqn:dS/dT}
\end{equation}
which is exactly the expression of the specific heat divided by temperature.

$S_{el}$ involves integral of the Landau level DOS and a term $- F(x)\ln{(F(x))} - (1-F(x))\ln{(1-F(x))}$. As shown in Supplementary Fig.\,\ref{fig:Entropy}a, this term exhibits a single-peak feature, in contrast to the double-peak feature that originated from $x^2(-dF(x)/dx)$ in $C_{el}/T$.
The calculated $S$ at various temperature is shown in Supplementary Fig.\,\ref{fig:Entropy}b. It is clear that $S$ shows a single peak structure when the Landau level crosses through the Fermi energy. However, the differential entropy $dS_{el}/dT \propto C_{el}/T$ is expected to show the double-peak structure. In Supplementary Fig.\,\ref{fig:Entropy}c we plot the calculated differential entropy, $(S(2\text{K})-S(1\text{K}))/1\text{K}$, and the expected double-peak feature is indeed present.

\begin{figure*}[th!]
  \renewcommand{\figurename}{Supplementary Figure} %  \thefigure {S\arabic{figure}}
  \centering
   \includegraphics[width= 0.9\linewidth]{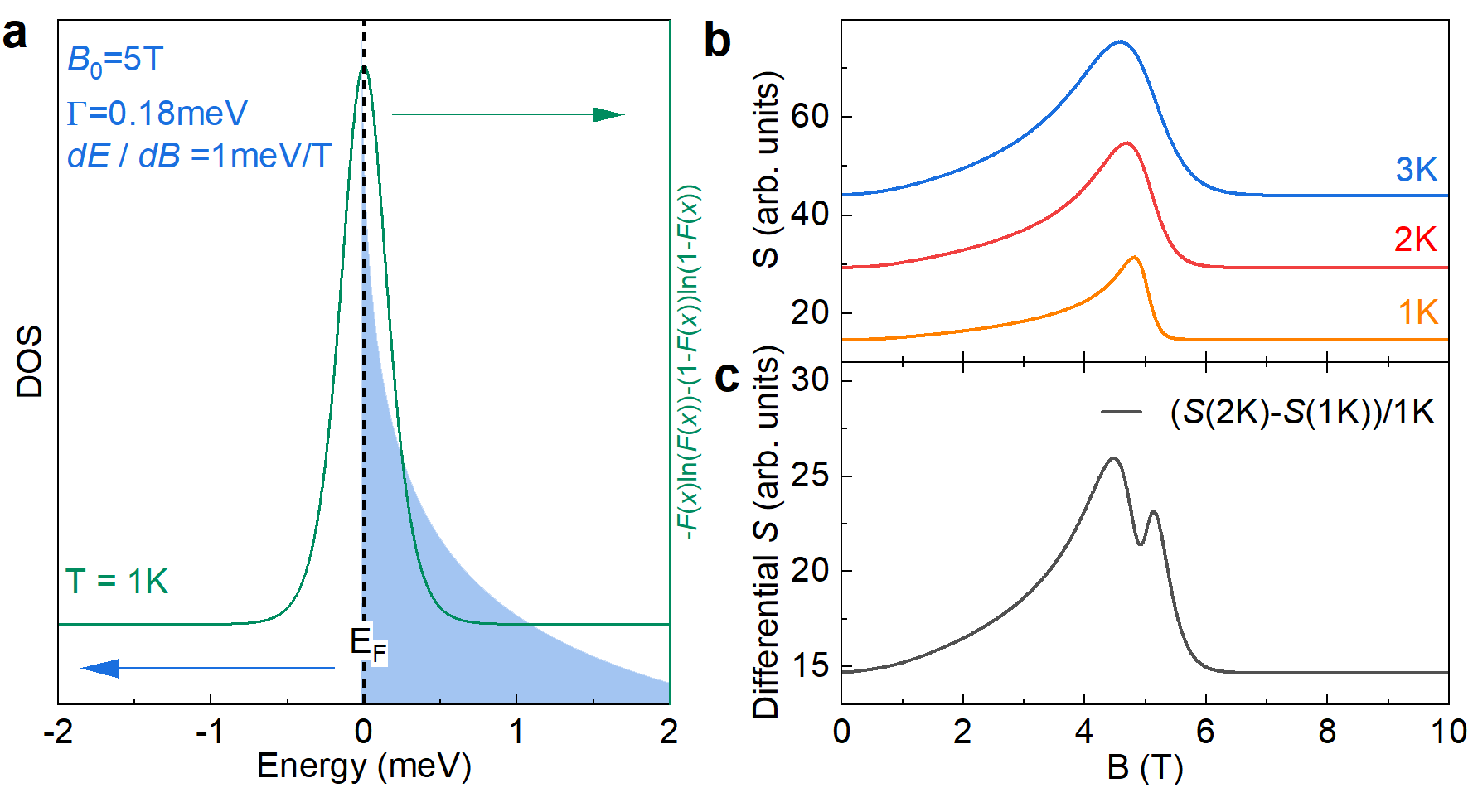}
  \caption{\textbf{a} Constructed Landau level DOS that used to calculate the entropy. \textbf{b} Calculated entropy $S$ ($\propto 1/T$ in MCE) for different temperatures in the vicinity of the crossing of Landau level with the Fermi energy. No double-peak structure is observed in the calculated entropy, in agreement with the absence of double-peak structure in MCE. \textbf{c} Calculated differential entropy in the vicinity of Landau level with Fermi energy. The double-peak structure is observed in the calculated differential entropy.}
  \label{fig:Entropy}
\end{figure*}

\clearpage

\begin{flushleft}
\section*{Supplementary Note 8: Maxima in the kernel function $-x^2 d$$F$$(x)/dx$ of specific heat}
\label{RosbifConst}  
\end{flushleft}

In the manuscript we state without justification that the maxima in $-x^2 dF(x)/dx$ occur at $x=\pm 2.4$. This value was determined by manually reading the peak position from a plot of the function. Here, for completeness, we attempt to derive this result using calculus. We define the Fermi-Dirac function $y=(1 + e^x)^{-1}$. Using the chain rule we can calculate the derivatives $dy/dx$ and $d^2y/dx^2$. We define the kernel function $z = -x^2 dy/dx$. We find the maxima in the kernel function by looking for zeros in the first derivative,
\begin{equation}
\renewcommand \theequation {\arabic{equation}}
\frac{dz}{dx} = -2 x \frac{dy}{dx} - x^2 \frac{d^2y}{dx^2} = x y^2 e^x (2 - 2xy e^x +x),     
\end{equation}
The maxima in $z$ correspond to the roots of the function
\begin{equation}
\renewcommand \theequation {\arabic{equation}}
0 =  (2 - 2xy e^x +x),     
\end{equation}
Multiply both sides by $y^{-1}$ and substituting $(1 + e^x)$ for $y^{-1}$ we obtain after simplification,
\begin{equation}
\renewcommand \theequation {\arabic{equation}}
0 =  (2+x) + (2-x) e^x,     
\end{equation}
Despite the apparent simplicity of this function we failed to find an analytical solution for the roots.  Instead we used Newton's method, to find the roots of a function f(x), using successive approximations, $x_{n+1} = x_n - f(x_n)/f'(x_n)$, using the initial guess $x_0=\pm 2$. This method converges rapidly to give the roots $x = \pm 2.399357280515468$, justifying \emph{a posteriori}, our approximation $x=\pm 2.4$.

\clearpage

\begin{flushleft}
\section*{Supplementary Note 9: Thermopower in graphite}
\label{Thermopower}   
\end{flushleft}

As mentioned in the main text, the negative and positive peaks in the thermopower have been experimentally observed. In this section, we demonstrate this feature using the thermopower data of graphite in the literature\,\cite{woollam1971graphite,zhu2010nernst}. As a step forward, we predict the splitting of the negative and positive peaks of thermopower to be $\simeq 3.09 k_BT$ based on the kernel term.

\subsection*{Supplementary Note 9.1: Negative and positive peaks in thermopower of graphite}
\label{NPinThermopower}
\begin{figure}[h!]
  \renewcommand{\figurename}{Supplementary Figure} %  \thefigure {S\arabic{figure}}
  \centering
  \includegraphics[width= 0.6\linewidth]{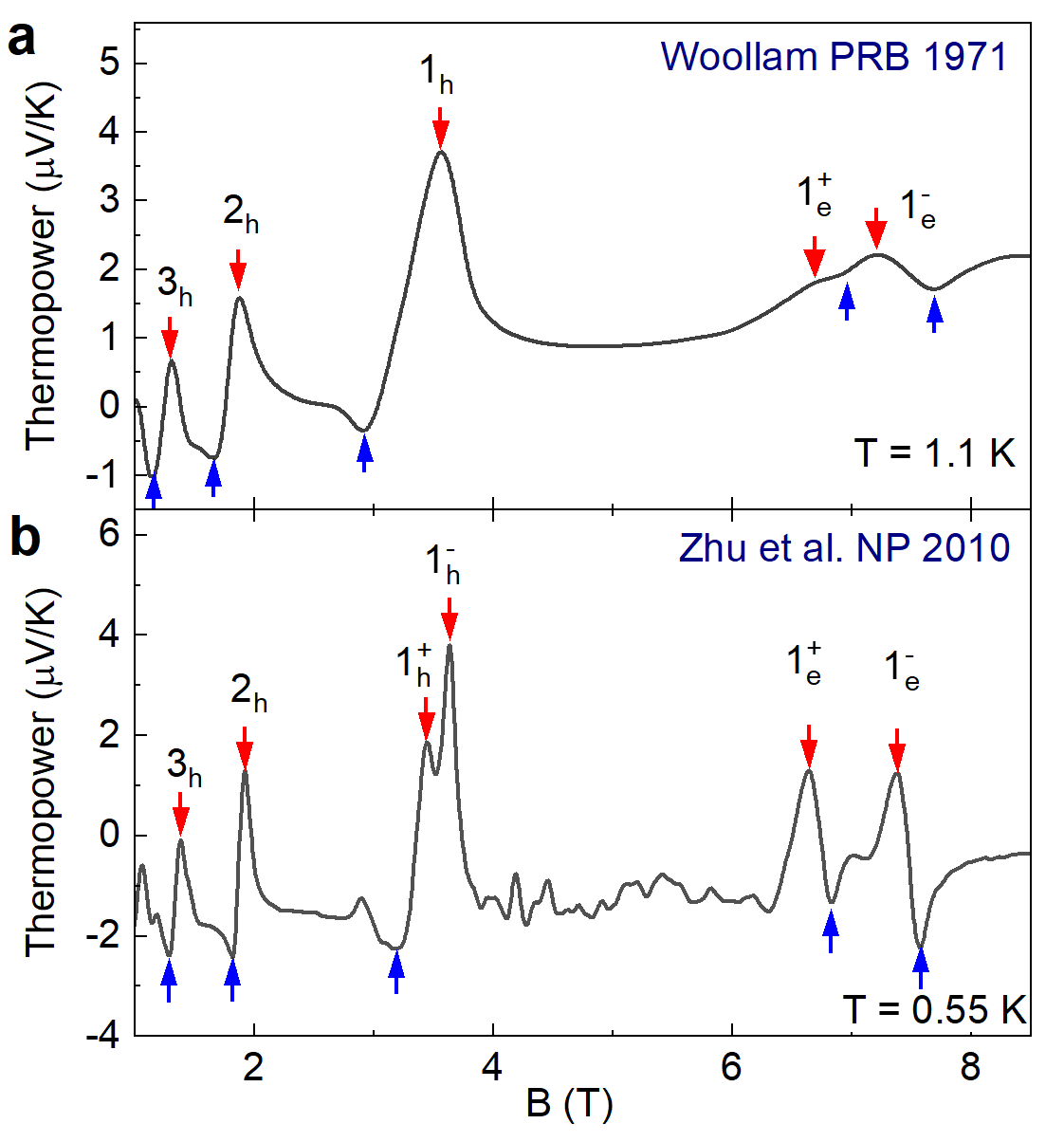}
      \caption{
      Thermopower as a function of magnetic field for graphite. The experimental data were digitalized from ref\,\cite{woollam1971graphite} \textbf{a} and ref\,\cite{zhu2010nernst} \textbf{b}. The blue and red arrows indicate the negative and postive peak position in the thermopower for a single DOS peak passing over the Fermi level.
      }
  \label{fig:MarkThermoP}
\end{figure}

Supplementary Fig.\,\ref{fig:MarkThermoP} shows the thermopower versus magnetic field for graphite. The experimental data were digitalized from ref\,\cite{woollam1971graphite,zhu2010nernst}. As seen in Supplementary Fig.\,\ref{fig:MarkThermoP}, the thermopower shows a negative and a positive peak for each Landau level passing over the Fermi energy, as marked by blue and red arrows.

It is important to note that this negative/positive peak feature are opposite for the electrons and holes. Namely, the positive peak locates at lower magnetic field position for the electrons, but locates at higher magnetic field position for the holes.

\subsection*{Supplementary Note 9.2: Predictions for thermopower}

\begin{figure*}[h!]
  \renewcommand{\figurename}{Supplementary Figure} %  \thefigure {S\arabic{figure}}
   \centering
   \includegraphics[width= 0.5\linewidth]{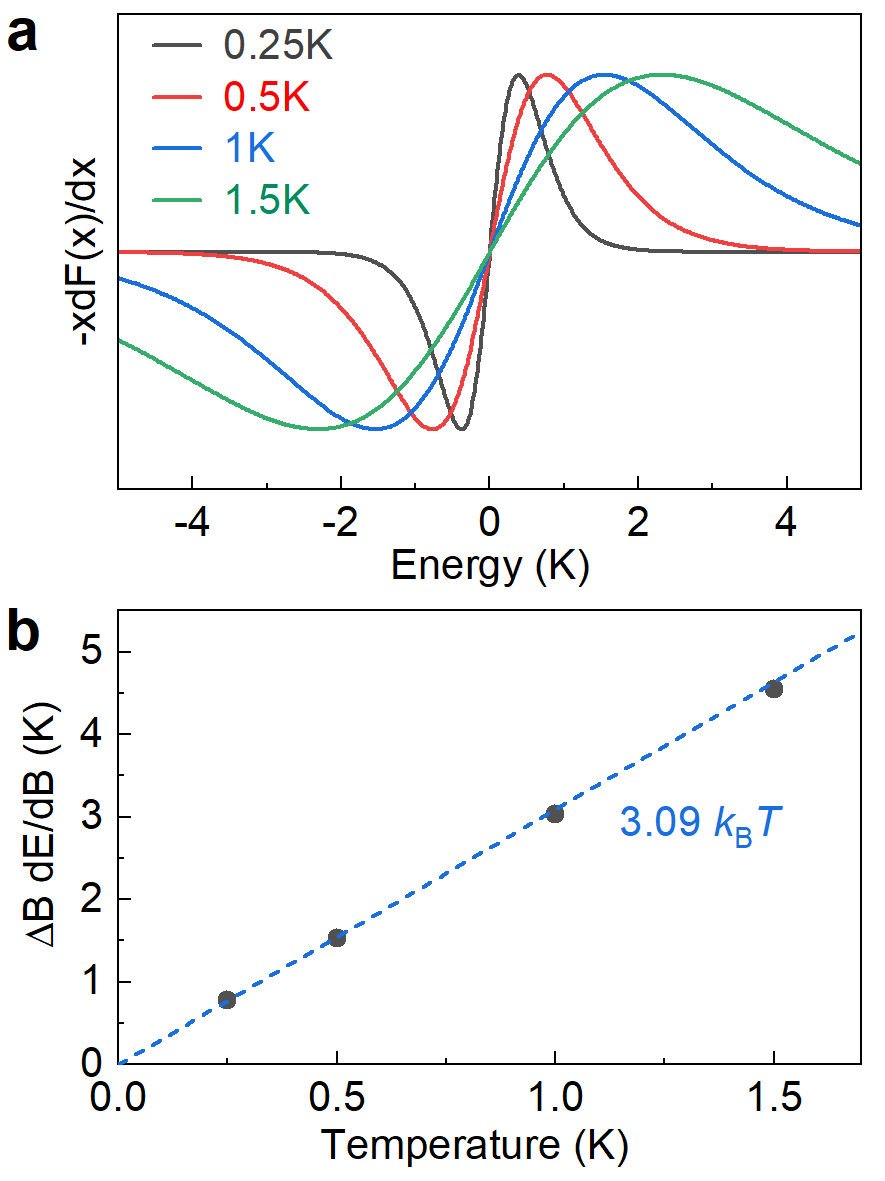}
  \caption{\textbf{a} The kernel term for thermopower $-x dF(x)/dx$ in the vicinity of the Fermi energy calculated at different temperatures. \textbf{b} The dashed line shows the theoretical $3.09k_BT$ splitting between the maximum and minimum of the kernel function $-x dF(x)/dx$. The symbols are the splitting manually read off from the plotted $-x dF(x)/dx$ in panel \textbf{a}.}
  \label{fig:Thermopower}
\end{figure*}

The kernel function for specific heat and thermal transport $x^2 dF/dx$, with $x=E/k_BT$, is an even function of energy around the Fermi energy, with maxima at $x = \pm 4.8$. With the exception of thermopower, other thermodynamic probes have a kernel function with a single maximum at $x=0$. The kernel function for thermopower $-x dF/dx$ is an odd function of energy around the Fermi energy, as can be seen in Supplementary Fig.\,\ref{fig:Thermopower}.

We define the kernel function $z = -x dy/dx$ with the Fermi-Dirac distribution function $y =(1 + e^x)^{-1}$. We find the maximum and minimum in the kernel function by looking for zeros in the first derivative,
\begin{equation}
\renewcommand \theequation {\arabic{equation}}
\frac{dz}{dx} = - \frac{dy}{dx} - x \frac{d^2y}{dx^2} =  y^3 e^x ((1+x) + (1-x) e^x),     
\end{equation}
The maximum and minimum in $z$ correspond to the roots of the function
\begin{equation}
\renewcommand \theequation {\arabic{equation}}
0 =  (1+x) + (1-x) e^x. 
\end{equation}

Using Newton's method the roots occur at $x = \pm 1.543404638418208$, so that the expected splitting of the maximum and minimum in thermopower, when a DOS singularity crosses the Fermi energy is $\simeq 3.09 k_BT$.

\clearpage
\begin{flushleft}
 \section*{Supplementary Note 10: Deviation from the expected $4.8 k_B T$ double-peak structure splitting in specific heat due to the asymmetry of the Landau level DOS}
\label{Deviation4p8}   
\end{flushleft}

The simple picture, in which maxima in $C_{el}/T \propto \int D(E) (-x^2 dF(x)/dx)$ occur when the DOS peak lies at the centre of the maxima in kernel term $-x^2 dF(x)/dx$ (located at $x=\pm 2.4$, $E = x k_B T$), has to be exact provided the DOS peak is symmetric (e.g. cusp-like DOS in Lifshitz transition). We validated this hypothesis by calculating the overlap integral versus energy, for different temperatures and width $\Gamma$, as the DOS peak passes through the Fermi energy. In Supplementary Fig.\,\ref{fig:GammaCorrection}a, we plot the double-peak structure splitting $\Delta B\,dE/dB$ in the simulated $C_{el}/T$ using a symmetric DOS peak, versus temperature. The dashed line is the expected variation if the splitting in $C_{el}/T$ exactly mimics the $4.8 k_B T$ splitting of $-x^2 dF(x)/dx$. In the inset of Fig.\ref{fig:GammaCorrection} (a), we plot the splitting in the calculated $C_{el}/T$ as a function of the Landau level width $\Gamma$. This is simply the slope of the $\Delta E$ versus $T$ plots in the main panel. As can be seen, the splitting in $C_{el}/T$ for a symmetric DOS peak are in good agreement with expected $4.8 k_B T$ regardless of the temperatures and the width $\Gamma$.

However, this is not exactly the case for a highly asymmetric DOS peak (e.g. DOS in Landau levels). The large asymmetric tail, on the high energy side of the `singularity' ($D(E)=\beta/(1+\sqrt{(E-E_0)/\Gamma})$) causes the peaks in $C_{el}/T$ to shift away from this condition. Although both peaks shift in the same direction (see Supplementary Fig.\,\ref{fig:SymUnsym}e), the shift of the peaks is not identical due to the asymmetric shape of DOS. For this reason, we performed similar simulation using a highly asymmetric DOS peak to evaluate how important are the deviations of the splitting in $C_{el}/T$ from the $4.8 k_B T$ splitting of the maxima in $-x^2 dF(x)/dx$.  In Supplementary Fig.\,\ref{fig:GammaCorrection}b we plot the double-peak structure splitting in the simulated $C_{el}/T$ using an asymmetric DOS peak, versus temperature. As can be seen, the splitting in $C_{el}/T$ is larger than $4.8 k_B T$ and the correction is of the order of 20\% for $\Gamma=0.2$\,meV.  When using the splitting in $C_{el}/T$ to determine for example the electronic $g$-factors using the coincidence method, it is important to use the splitting which corresponds to the correct width of the Landau level, which can be extracted from the fit to the experimental $C_{el}/T$ versus $B$ data.

\begin{figure*}[b]
  \renewcommand{\figurename}{Supplementary Figure} %  \thefigure {S\arabic{figure}}
  \centering
   \includegraphics[width= 0.9\linewidth]{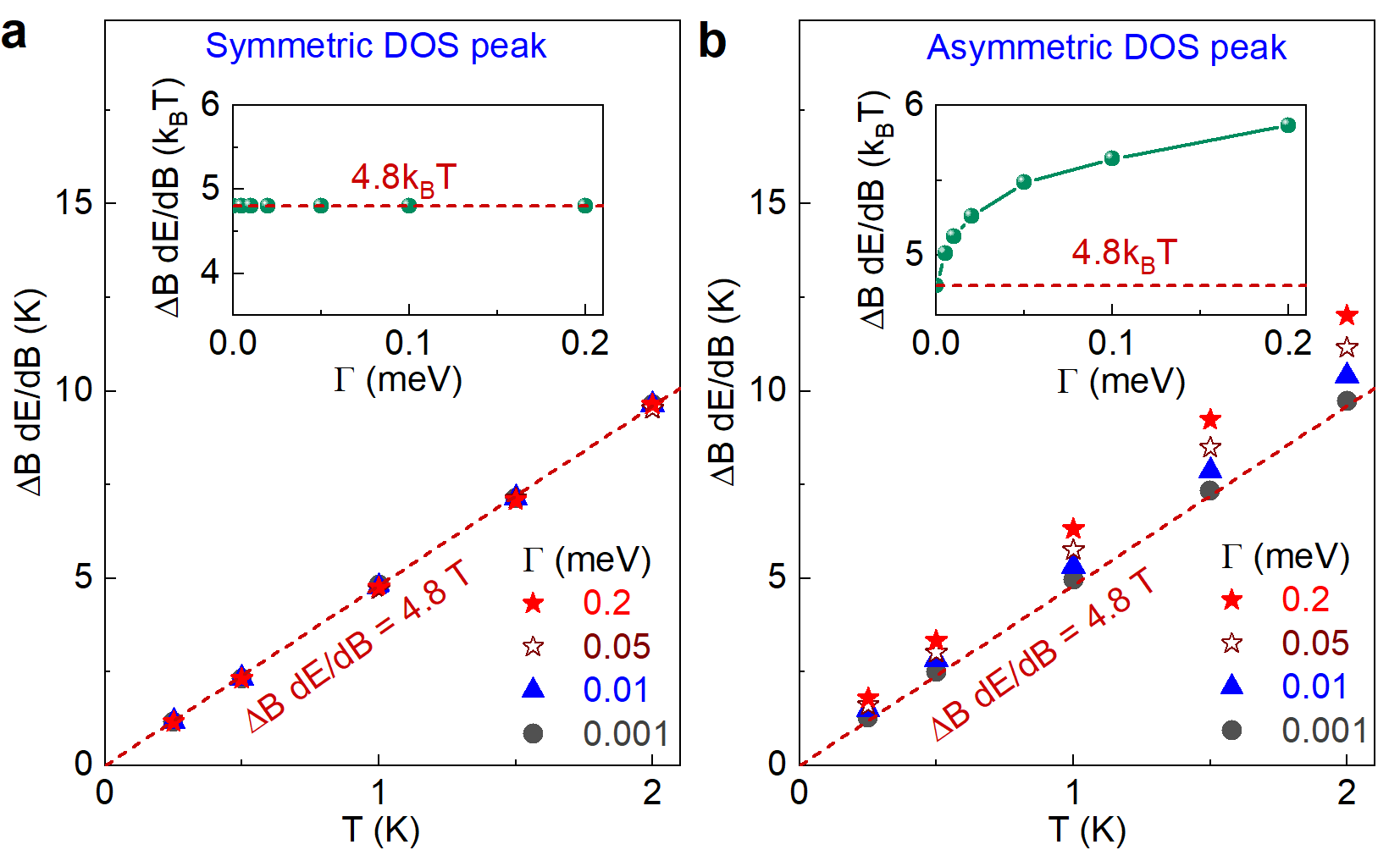}
  \caption{
  Calculated double-peak structure splitting in the specific heat $\Delta B\,dE/dB$ in natural units of Kelvin as a function of temperature, for selected FWHM $\Gamma$ of \textbf{a} symmetric and \textbf{b}\,asymmetric DOS peak. The red dashed line is the naively expected $\Delta B\,dE/dB = 4.8 T$ dependence. The inset show the double-peak structure splitting in the calculated specific heat as a function of DOS peak width $\Gamma$.}
  \label{fig:GammaCorrection}
\end{figure*}

\clearpage

\begin{flushleft}
\section*{Supplementary Note 11: Peak position in MCE and specific heat for symmetric and asymmetric DOS}
\label{SI:MCEdeviation}   
\end{flushleft}

\begin{figure}[h!]
  \renewcommand{\figurename}{Supplementary Figure} %  \thefigure {S\arabic{figure}}
  \centering
  \includegraphics[width= 0.8\linewidth]{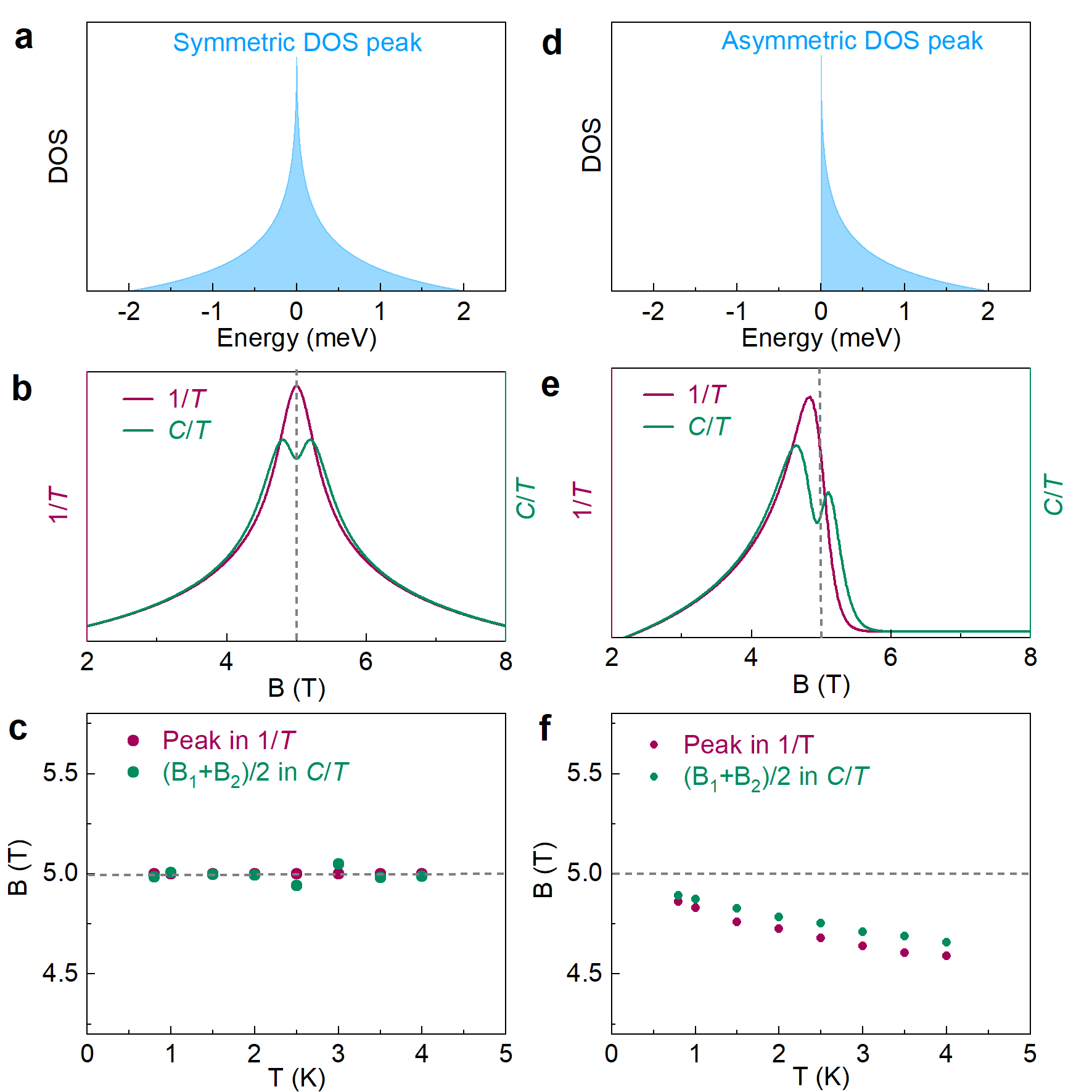}
      \caption{\textbf{a - b} Symmetric and asymmetric DOS peak used to calculate the MCE ($1/T$) and the specific heat ($C/T$). In each case the DOS peak crosses $E_\text{F}$ at $B$ = 5\,T. \textbf{c - d} Calculated $1/T$ and $C/T$ curve at $T$ = 1\,K for symmetric and asymmetric DOS peaks, respectively. \textbf{d - e} Position of the peak in $1/T$ and the center of double-peak structure (($B_1+B_2)/2$) at different temperature for symmetric and asymmetric DOS, respectively.}
  \label{fig:SymUnsym}
\end{figure}

In a simple picture, with a symmetric DOS peak, the peak position of MCE ($1/T$) lies exactly in the middle of the double peaks in $C_\text{el}/T$ (see Supplementary Fig.\,\ref{fig:SymUnsym}a-c). However, when the DOS peak is asymmetric, the peak position of MCE ($1/T$) deviates from the center of double peaks in $C_\text{el}/T$ (see Supplementary Fig.\,\ref{fig:SymUnsym}d-f). For an asymmetric DOS, both features occur at magnetic fields slight below the $B=5$\,T crossing of the Fermi energy.

\clearpage

\begin{flushleft}
\section*{Supplementary Note 12: Double-peak structure of specific heat near the Lifshitz transition}
\label{SI:Lifshitz}  
\end{flushleft}

\begin{figure}[h!]
  \renewcommand{\figurename}{Supplementary Figure} %  \thefigure {S\arabic{figure}}
  \centering
  \includegraphics[width= 0.9\linewidth]{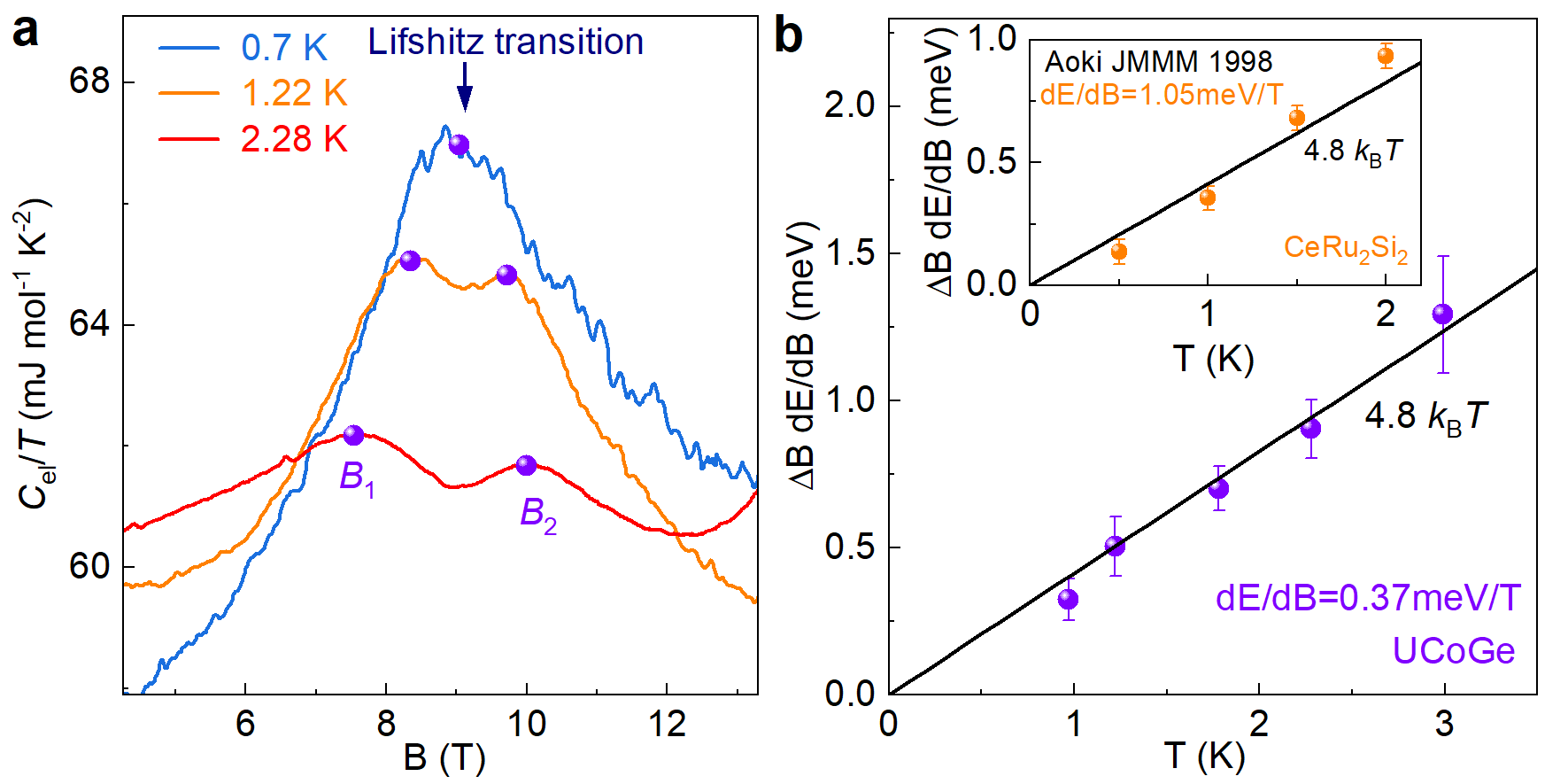}
      \caption{\textbf{a} Fermionic quasiparticle specific heat divided by temperature $C_{\text{el}}/T$ in UCoGe as a function of magnetic field in the vicinity of Lifshitz transition showing a double-peak structure splitting. Note that the symmetrical nature of the peaks, reflects the cusp like ``singularity'' in the DOS close to the Lifshitz transition, compared to the asymmetric nature of the Landau level DOS ``singularity'' in graphite. \textbf{b} Temperature dependence of the energy $\Delta B\,dE/dB$ through which ``singularity'' moves in UCoGe. The inset shows the same plot for $\Delta B$ extracted from the published $C_{\text{el}}/T$ data of Aoki \emph{et al.} \cite{aoki1998thermal} on the heavy fermion compound CeRu$_2$Si$_2$.}
  \label{fig:UCoGe}
\end{figure}

In addition to the quantum oscillation in graphite reported here, a double-peak structure (double peak) is occasionally observed in quasi-particle specific heat $C_{\text{el}}/T$ near the Lifshitz transition, where a  cusp like “singularity" occurs in the DOS\,\cite{mori2019controlling,miyake2006true}. As an example, in Supplementary Fig.\,\ref{fig:UCoGe}a we show $C_{\text{el}}/T$ of UCoGe in the vicinity of Lifshitz transition. UCoGe is ferromagnetic superconductor that exhibiting a Lifshitz transition at $B_c$=9.5\,T\,\cite{bastien2016lifshitz}. The field sweep $C_{\text{el}}/T$ near the Lifshiz transition of UCoGe exhibits same double-peak structure observed in the specific heat of graphite. A similar behaviour is also observed in the Lifshitz transition of CeRu$_2$Si$_2$ at $B_c$=7.7\,T \,\cite{aoki1998thermal}. The $dE/dB$ for both CeRu$_2$Si$_2$ and UCoGe is estimated from the relation $\Delta B(dE/dB) = 4.8 k_BT$, as shown in Supplementary Fig.\,\ref{fig:UCoGe}b.

\clearpage

\begin{flushleft}
\section*{Supplementary Note 13: Fitting of $C/T$ near the Lifshitz transition of CeRu$_2$Si$_2$}
\label{LT_CeRu2Si2}   
\end{flushleft}

\begin{figure}[h!]
  \renewcommand{\figurename}{Supplementary Figure} %  \thefigure {S\arabic{figure}}
  \centering
  \includegraphics[width= 0.95\linewidth]{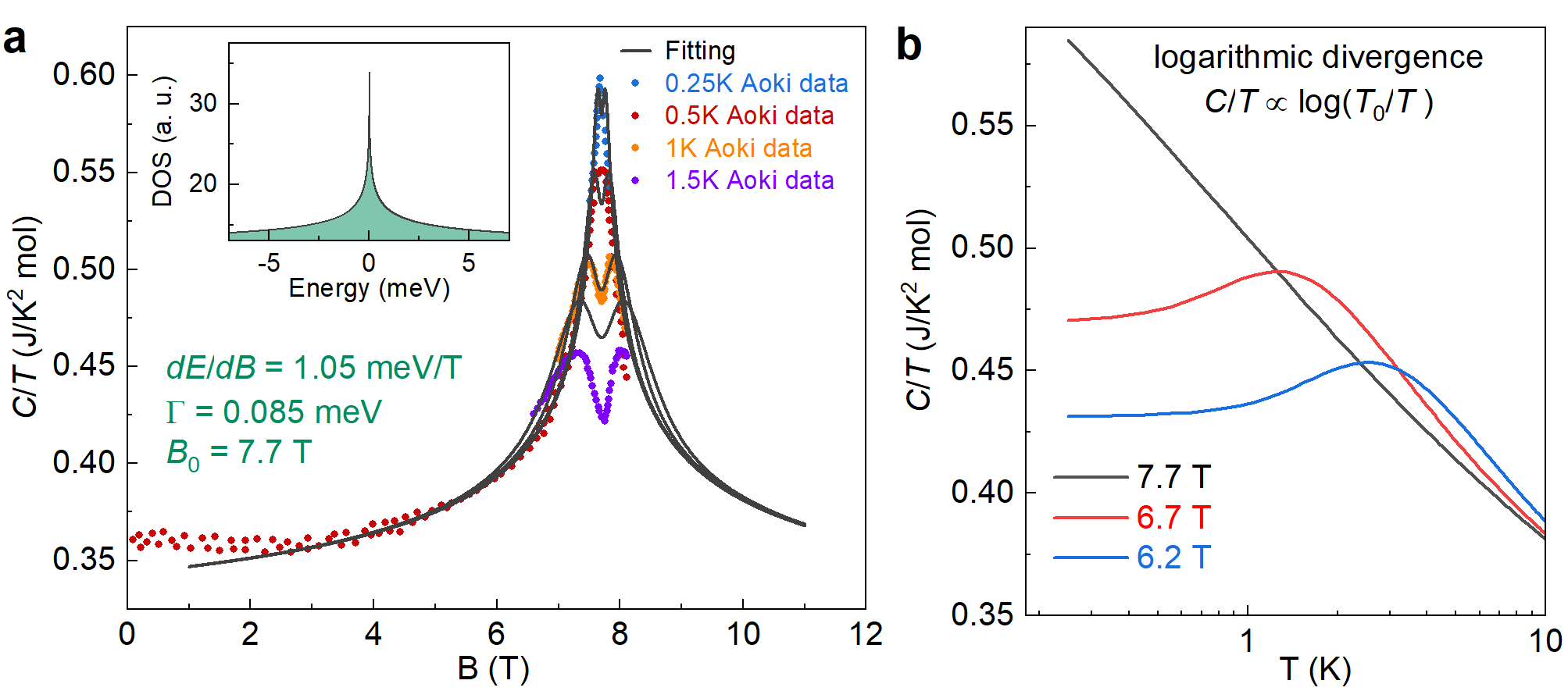}
      \caption{\textbf{a} Data points are the digitized $C/T$ versus $B$ data near the Lifshitz transition of CeRu$_2$Si$_2$ from ref\,\cite{aoki1998thermal}. The solid black lines is the calculated behavior of $C/T$, for a \emph{single} symmetric DOS peak passing through the Fermi energy, using the parameters indicated on the figure. The inset shows the \emph{single} symmetric DOS peak used in the calculation which nicely reproduces the double-peak structure in $C/T$ versus $B$ lending further support to our model. \textbf{b}\,Semi-log plot of calculated $C/T$ versus $T$ using a \textit{single} symmetric DOS peak at constant magnetic fields. At critical field $B$=7.7\,T, $C/T$ versus $T$ shows a logarithmic divergence.}
  \label{fig:FitCeRu2Si2}
\end{figure}

Supplementary Fig.\,\ref{fig:FitCeRu2Si2}a shows the digitized $C/T$ data points near the Lifshitz transition of CeRu$_2$Si$_2$ from Aoki et al.\,\cite{aoki1998thermal}. It is predicted theoretically that the DOS peak in the subband edge of CeRu$_2$Si$_2$ is cusp-like ($D(E) \propto E^{-0.5}$)\,\cite{miyake2006true}. We therefore use a symmetric single cusp-like DOS peak (see inset of Supplementary Fig.\,\ref{fig:FitCeRu2Si2}a) to calculate the $C/T$, as shown by black solid lines in Supplementary Fig.\,\ref{fig:FitCeRu2Si2}a, using the parameters indicated on the figure. The fits are in an excellent agreement with the experimental data up to 1\,K, but start to deviate in overall amplitude when the temperature is higher than 1\,K. The discrepancy occurs because the temperature-dependent of the shape of DOS peak is not taken into account in our simple model. 

The excellent agreement between the calculated $C/T$ using a single symmetric DOS peak and the experimental results clearly demonstrates that only one subband (DOS peak) passes through the Fermi energy at the Lifshitz transition of CeRu$_2$Si$_2$. Moreover, the MCE curve of CeRu$_2$Si$_2$ only exhibit a single-peak feature\,\cite{aoki1998thermal}, which is fully consistent with our model. The simultaneous occurrence of single- and double-peak features in MCE and specific heat are smoking gun thermodynamic signatures of a single DOS peak crossing the Fermi level. Moreover, the magneto-resistance, Hall resistivity and thermopower measurements on CeRu$_2$Si$_2$ also suggest the same scenario\,\cite{Daou2006PRL}\cite{Pfau2012PRB}, that is, a single spin subband crosses the Fermi energy at the Lifshitz transition of CeRu$_2$Si$_2$.

In heavy fermion system, it is known that a 'logarithmic divergence' feature in temperature sweep of specific heat ($C/T \propto \text{log}\,(T_0/T)$) is a signature of non-Fermi Liquid behaviour \,\cite{Steglich2004PRL}. Here, it is interesting to note that such logarithmic divergence feature can also be observed in $C/T$ versus $T$ for a fermionic singularity in DOS that passes over the Fermi energy. In Supplementary Fig.\,\ref{fig:FitCeRu2Si2}b, we show the semi-log plot of calculated $C/T$ as a function of temperature using \textit{single} symmetric DOS peak at constant magnetic fields. If the system has a logarithmic divergence behaviour, the temperature-sweep of $C/T$ exhibits a straight line in the semi-log plot. As seen in Supplementary Fig.\,\ref{fig:FitCeRu2Si2}b, $C/T$ versus $T$ exhibits as a straight line at the critical field $B$ = 7.7\,T. Therefore, the logarithmic divergence in $C/T$ is not only due to an existence of a canonical quantum critical point, but can also be attributed to the formation of a fermionic DOS singularity.

\clearpage

\begin{flushleft}
\section*{Supplementary Note 14: Advantage of extracting effective $g$-factor from double-peak structure in $C_{el}/T$}
\label{SecAdvgfactor}    
\end{flushleft}

In most cases of quantum oscillations, the exact shape of the DOS is unknown which makes it difficult fit the data in order to extract the $g$-factor, $dE/dB$ \emph{etc}. Typically the $1/B$ periodicity of the oscillations are used to exactly calipers the Fermi surface.

As described in more detail below, to extract the $g$-factor using techniques such as SdHs, dHvA, MCE \emph{etc}, one has to know the Landau index (orbital quantum number) for each peak, and the system dependent Fermi energy shift. This type of difficulty is discussed at some length in the classic book of D. Shoenberg $\ll$ Magnetic Oscillations in Metals$\gg$\,\cite{shoenberg2009magnetic}.

We stress once more, that crucially, the double-peak feature observed in specific heat (or eventually thermal conductance), when a single DOS peak crosses the Fermi energy, allows us to estimate the $g$-factor, without having to make any assumptions concerning the Landau index or Fermi energy shift.

\subsection*{Supplementary Note 14.1: Extraction of g-factor from position of spin-up and spin-down peaks in SdH, dHvA, MCE}

Quantum oscillations, which are driven by the magnetic field dependent Landau level degeneracy, exactly caliper the Fermi surface, independently of the values of the cyclotron, or Zeeman energies. They are therefore, by definition, not well adapted to determine these quantities. In SdH, dHvA MCE measurements a spin split Landau level crossing the Fermi energy gives rise to two peaks. We can define a magnetic field splitting $\Delta B = B^+ - B^-$, where $B^+$ and $B^-$ are the magnetic field position of spin up and spin down features. To a reasonable approximation, the $g$-factor can be estimated using,

\begin{equation}
\renewcommand \theequation {\arabic{equation}}
\Delta B = \frac{g^* \mu_B B_m}{(N+1/2)\hbar e/m^* - S_F},
\label{Eq.g-single}
\end{equation}

where $B_m = (B^+ + B^-)/2$ is the mean field position for $B^+$ and $B^-$, $N$ is the Landau index, $m^*$ is the effective mass, $S_F$ is the slope of the Fermi energy in the $N$th Landau level. To obtain g-factor from Supplementary Eq.\,(\ref{Eq.g-single}), the Landau index and Fermi energy shift for the relevant spin-split Landau level are required, which are generally difficult to identify in a new system. 

\subsection*{Supplementary Note 14.2: Extraction of $g$-factor from double peak in $C/T$ using coincidence method}

\begin{figure}[h!]
  \renewcommand{\figurename}{Supplementary Figure} %  \thefigure {S\arabic{figure}}
  \centering
  \includegraphics[width= 0.45\linewidth]{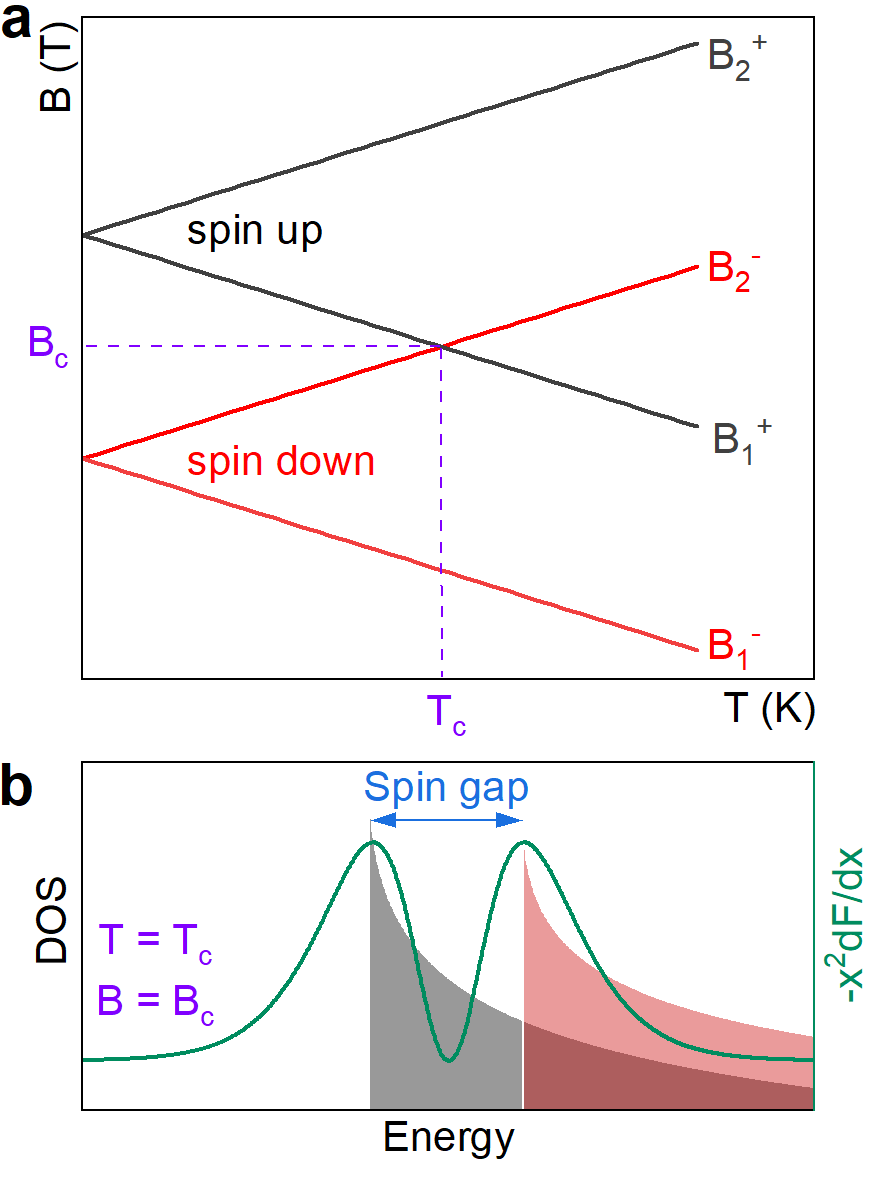}
      \caption{\textbf{a} Schematic to show the magnetic field position of the two independent double-peak structures $B_{1,2}^+$, $B_{1,2}^-$ as a function of temperature when spin up/spin down levels with the same orbital quantum number are in the vicinity of the Fermi energy. \textbf{b} Schematic showing the coincidence condition at $T=T_c$ when the spin-up/down spin split levels simultaneously lie at the centre of one of the two maxima in $-x^2 dF/dx$}.
  \label{fig:Advgfactor}
\end{figure}

In specific heat (eventually thermal transport) measurements, a spin split Landau level crossing the Fermi energy should gives rise to four peaks \emph{i.e.} two independent double-peak structure. However, under special conditions referred to as coincidence, two of the peaks occur at exactly the same magnetic field.  Supplementary Fig.\,\ref{fig:Advgfactor}a shows the magnetic field position of the quadruple-peak structure $B_{1,2}^+$, $B_{1,2}^-$ as a function of temperature for a spin up and a spin down levels with same Landau index $N$. At a critical temperature $T_c$, the $B_{2}^-$ peak and the $B_{1}^+$ peak occur at the same magnetic field $B_c$. At this temperature, the $4.8k_B T_c$ splitting of $-x^2dF/dx$ has the correct value, so that the spin-up and spin-down DOS peaks are simultaneously located at one of the two maxima of $-x^2dF/dx$, as schematically illustrated in Supplementary Fig.\,\ref{fig:Advgfactor}b.

At the experimentally determined $T=T_c$, the energy gap between spin-up and spin-down levels is equal to the splitting of the maxima in $-x^2dF/dx$, therefore, we have,

\begin{equation}
\renewcommand \theequation {\arabic{equation}}
g^* \mu_B B_c = 4.8 k_BT_c,
\label{Eqn:g-double}
\end{equation}
Therefore, the coincidence condition (experimentally determined $B_c$ and $T_c$) allows us to extract the g-factor without knowing the Landau index, or making any assumptions concerning the Fermi energy shift.

\clearpage

\bibliography{Manuscript.bib}% Produces the bibliography via BibTeX.

%\end{document}
%
% ****** End of file apssamp.tex 

\end{document}